\documentclass[11pt]{article}
\pdfoutput=1 

\usepackage{cancel,slashed}
\usepackage{bbm}
\usepackage{amssymb}
\usepackage{amsfonts}
\usepackage{amsmath}
\usepackage{graphicx}
\usepackage{cite}
\usepackage{bbold}
\usepackage{latexsym}
\usepackage{color}
\usepackage{xypic}
\usepackage{cancel,slashed}
\usepackage[linktocpage]{hyperref}
\usepackage{color}
\usepackage{transparent}
\usepackage{footmisc}
\usepackage{mathtools}
\usepackage{array}
\usepackage{makecell}
\usepackage[utf8]{inputenc}

\makeatletter
\def\@xfootnote[#1]{%
  \protected@xdef\@thefnmark{#1}%
  \@footnotemark\@footnotetext}
\makeatother

\newcommand{\bmat}{\left(\begin{array}}
\newcommand{\emat}{\end{array}\right)}

\def\-{\hphantom{-}}

\def\s2{\frac{1}{\sqrt2}}

\def\beq{\begin{equation}}
\def\eeq{\end{equation}}
\def\beqa{\begin{eqnarray}}
\def\eeqa{\end{eqnarray}}

\def\Tr{{\rm Tr \,}}
\def\dim{{\rm dim \,}}

\def\Dsl{\,\raise.15ex\hbox{/}\mkern-13.5mu D} 

\def\be{\begin{equation}}
\def\ee{\end{equation}}
\def\bea{\begin{eqnarray}}
\def\eea{\end{eqnarray}}

\def\cN{\mathcal{N}}

\def\sm2{{\mbox{\small 2}}}

\newcommand{\bp}{\begin{pmatrix*}[r]}  
\newcommand{\ep}{\end{pmatrix*}}  
\newcommand{\bpp}{\begin{pmatrix}}  
\newcommand{\epp}{\end{pmatrix}}  
\newcommand{\bcd}{\begin{center}
\begin{tikzcd}}
\newcommand{\ecd}{\end{tikzcd} \end{center}}

\def\1{\mathbb{1}}

\usepackage{tikz}
\usetikzlibrary{arrows,snakes,backgrounds}
\usetikzlibrary{shapes.misc}
\usetikzlibrary{decorations.pathreplacing}

\tikzstyle{gauge}=[circle,draw=blue!50,fill=blue!20,thick, inner sep=0pt,minimum size=1cm]
\tikzstyle{circ}=[circle,draw,thick,
inner sep=0pt,minimum size=1cm]
\tikzstyle{Dfive}=[circle, cross, draw=black!50,thick, inner sep=0pt,minimum size=0.4cm]
\tikzstyle{DfiveBig}=[circle, cross, draw=black!50,thick, inner sep=0pt,minimum size=0.6cm]
\tikzstyle{lpre}=[-,shorten <=0pt,>=stealth', very thick]
\tikzstyle{lpost}=[-,shorten >=0pt,>=stealth',very thick]
\tikzstyle{global}=[rectangle,draw=black!50,fill=black!20,thick,
inner sep=0pt,minimum size=1cm]
\tikzstyle{pre}=[<-,shorten <=0pt,>=stealth', very thick]
\tikzstyle{post}=[->,shorten >=0pt,>=stealth',very thick]
\tikzstyle{bpost}=[->, shorten >=2pt, shorten <=2pt, >=stealth', very thick]
\tikzstyle{bpre}=[<-, shorten >=2pt, shorten <=2pt, >=stealth', very thick]
\tikzstyle{nodo}=[circle,draw=black,fill=black,thick, inner sep=0pt,minimum size=2mm]
\tikzstyle{nodoblu}=[circle,draw=blue,fill=blue,thick, inner sep=0pt,minimum size=2mm]
\tikzset{cross/.style={path picture={ 
			\draw[black]
			(path picture bounding box.south east) -- (path picture bounding box.north west) (path picture bounding box.south west) -- (path picture bounding box.north east);
		}}}
		
		\tikzset{
			partial ellipse/.style args={#1:#2:#3}{
				insert path={+ (#1:#3) arc (#1:#2:#3)}
			}
		}

\begin{document}
\pagestyle{plain}

\makeatletter
\@addtoreset{equation}{section}
\makeatother
\renewcommand{\theequation}{\thesection.\arabic{equation}}
\pagestyle{empty}
\rightline{ROM2F/2018/05}
\rightline{IFT-UAM/CSIC-18-93}
\vspace{0.5cm}
\begin{center}
\Huge{{SUSY enhancement from T-branes}
\\[15mm]}
\normalsize{Federico Carta,$^1$\footnote[$\dagger$]{La Caixa-Severo Ochoa Scholar} Simone Giacomelli,$^{2,3}$ and Raffaele Savelli$^4$ \\[10mm]}
\small{
${}^1$ \it Departamento de F\'isica Te\'orica and Instituto de F\'{\i}sica Te\'orica UAM-CSIC, \\Universidad Aut\'onoma de Madrid, Cantoblanco, 28049 Madrid, Spain\\[2mm] 
${}^2$ International Center for Theoretical Physics, Strada Costiera 11, 34151 Trieste, ITALY.\\[2mm]
${}^3$ INFN, Sezione di Trieste, Via Valerio 2, 34127 Trieste, ITALY\\[2mm] 
${}^4$ Dipartimento di Fisica, Universit\`a di Roma ``Tor Vergata'' \& INFN - Sezione di Roma2 \\ Via della Ricerca Scientifica, I-00133 Roma, ITALY
\\[10mm]} 
\normalsize{\bf Abstract} \\[8mm]
\end{center}
\begin{center}
\begin{minipage}[h]{15.0cm} 

We use the F-theoretic engineering of four-dimensional rank-one superconformal field theories to provide a geometric understanding of the phenomenon of supersymmetry enhancement along the RG flow, recently observed by Maruyoshi and Song. In this context, the superpotential deformations responsible for such flows are interpreted as T-brane backgrounds and encoded in the geometry of elliptically-fibered fourfolds. We formulate a simple algebraic criterion to select all supersymmetry-enhancing flows and, without any maximization process,  derive the main features of the corresponding $\cN=2$ theories in the infrared.

\end{minipage}
\end{center}
\newpage
\setcounter{page}{1}
\pagestyle{plain}
\renewcommand{\thefootnote}{\arabic{footnote}}
\setcounter{footnote}{0}


\tableofcontents


\section{Introduction}
\label{s:intro}

A few years ago, Maruyoshi and Song \cite{Maruyoshi:2016tqk, Maruyoshi:2016aim} provided convincing evidence for a remarkable phenomenon which takes place along certain RG flows of four-dimensional (4d) supersymmetric field theories. Their starting point is a non-necessarily Lagrangian 4d $\cN=2$ superconformal field theory with a non-Abelian flavor symmetry. To this theory they add a chiral field $M$ transforming in the adjoint of the flavor group and the following superpotential deformation
\be\label{DefIntro}
\delta W =\Tr(\mu M)\,,
\ee
where $\mu$ is the moment map associated with the flavor symmetry. The vacuum expectation value of $M$ is taken to be nilpotent, and hence the above deformation only preserve a $\cN=1$ subalgebra of the original $\cN=2$ algebra. For most of the choices of nilpotent orbit, the RG flow triggered by this deformation has a $\cN=1$ fixed point in the infrared (IR), but for some specific choices a supersymmetry enhancement occurs at low energy, giving back a $\cN=2$ superconformal theory in the IR. The latter theory, however, is never the same as the starting one, and in all known cases it is non-Lagrangian.

Besides being interesting in its own right, this phenomenon turns out to be very useful as it can provide candidate ultraviolet (UV) Lagrangians for intrinsically strongly-coupled field theories, like the Argyres-Douglas theories \cite{Argyres:1995jj,Eguchi:1996ds}, which are typically the IR fixed points of such flows. After this discovery, significant effort has been made in studying more general sets of theories and of flows that may have the same property \cite{Agarwal:2016pjo, Agarwal:2017roi, Benvenuti:2017bpg, Giacomelli:2017ckh, Maruyoshi:2018nod}, and also in finding general criteria for the appearance of supersymmetry enhancement \cite{Evtikhiev:2017heo, Giacomelli:2018ziv}.

Despite many details and the systematics are now clearer, however, the very reason why this phenomenon takes place remains fairly obscure. This is mainly due to the fact that so far supersymmetry enhancement has been argued for on the basis of an iterative a-maximization process \cite{Intriligator:2003jj}, used to derive the correct IR R-charge of the operators that remain coupled to the theory. 

The aim of this paper is to shed light on the mechanism at the origin of this peculiar phenomenon, by analyzing it from a geometric point of view. We focus on 3d/4d field theories of rank $1$, and engineer them in the context of M/F-theory, as theories on a M2/D3-brane probing isolated singularities of elliptically fibered ALE spaces.\footnote{See the recent paper \cite{Apruzzi:2018xkw}, where the context of F-theory is adopted to engineer analogous types of 4d deformations. Differently from the present work, however, RG flows are analyzed using standard field-theory techniques.} In this geometric set-up, the deformations \eqref{DefIntro} with a nilpotent vev for the adjoint chiral show up as ``T-brane'' deformations, much in the same spirit of \cite{Heckman:2010qv}. T-branes \cite{Cecotti:2010bp} can be seen as bound states of ordinary D-branes, characterized by non-commuting vev's for two of their worldvolume scalars.\footnote{Further readings on the subject of T-branes include  \cite{Donagi:2003hh,Hayashi:2009bt,Chiou:2011js,Donagi:2011jy,Donagi:2011dv,Font:2013ida,Anderson:2013rka,DelZotto:2014hpa,Collinucci:2014taa,Collinucci:2014qfa,Marchesano:2015dfa,Cicoli:2015ylx,Carta:2015eoh,Collinucci:2016hpz,Bena:2016oqr,Marchesano:2016cqg,Mekareeya:2016yal,Ashfaque:2017iog,Anderson:2017rpr,Bena:2017jhm,Collinucci:2017bwv,Cicoli:2017shd,Marchesano:2017kke}.} In the present context, we consider T-branes made from stacks of non-perturbative 7-branes (or of ordinary D6-branes for the 3d case). The latter play the role of flavor branes from the probe viewpoint, and we interpret $M$ as the vev of the background worldvolume complex scalar of the 7-brane stack.

Despite $\langle M\rangle$ being nilpotent, fluctuations make $M$ non-nilpotent. This allows us to read the effects of the deformations at the level of geometry, in the M/F-theory lift of the probed T-brane configurations. This is in contrast to T-branes of the nilpotent type, which are completely invisible to the M/F-theory geometry.\footnote{See \cite{Collinucci:2016hpz,Collinucci:2017bwv} for a probe analysis of nilpotent T-branes.} It is precisely by looking at the behavior of this geometry along the RG flow that we will be able to conclude whether supersymmetry will enhance or not in the IR.

Our logic can be summarized as follows for the 4d case (the 3d one is analogous): We start with the local geometry of a Calabi-Yau twofold in F-theory. The deformation is then encoded in the hypersurface equation for a Calabi-Yau fourfold, which characterizes the flow of the $\cN=1$ theory between the fixed points. If supersymmetry is to enhance at the IR fixed point, the fourfold must factorize into a Calabi-Yau twofold and a trivial factor. By treating the various hypersurfaces homogeneously, we will be able to determine in which cases this factorization (and thus the enhancement) occurs, and in which cases instead the probed IR geometry is higher-dimensional, giving rise to 4d theories with only four supercharges. Finally, for the cases that display supersymmetry enhancement, we will find the candidate IR $\cN=2$ theory together with the correct conformal dimension of its Coulomb-branch operator, by reading off the corresponding probed geometry.

Our results are in perfect agreement with those known from the field-theory analysis. Remarkably, we do not make use of any maximization procedure in our study, but only perform algebraic manipulations to derive the relevant IR quantities.

The paper is organized as follows: In Section \ref{sec:3d} we discuss the case of Abelian theories in three dimensions and their realization in M-theory. This part also serves to illustrate our approach and how the infrared effective geometry can be analyzed. In Section \ref{sec:4d} we turn our attention to four dimensional theories living on D3-branes probing F-theory singularities. We derive the conditions one has to impose in order to have supersymmetry enhancement and describe explicitly in four non-trivial cases how to determine the low-energy theory from the underlying geometry. We discuss three cases of infrared enhancement and one case in which supersymmetry does not enhance. In Appendix \ref{Scan} we collect some basic properties of the RG flows of interest for us, and provide an alternative derivation of the criterion for enhancement based on the Seiberg-Witten geometry alone. This is then applied to perform a detailed scan of all the nilpotent orbits for rank-$1$ theories engineered in F-theory. Only orbits for which enhancement occurs pass our criterion.

\section{Warm-up: SQED in three dimensions}
\label{sec:3d}

In this section we would like to introduce the main ideas behind our investigation. We do it in the context of 3d Abelian field theories because, on the one hand the analysis is technically easier, and on the other hand this allows us to lay down the general string-theory set-up which, with few important variations, will also be relevant for the study of 4d theories.

\subsection{Geometric set-up}\label{ss:geo3d}

The theory we start from is engineered by type IIA string theory with a single D2-brane probing a stack of $N$ D6-branes in flat space-time. Branes are extended as in the following table:

\begin{center}
$
\begin{array}{c|cccccccccc}
{\rm Type \; IIA}& 0 & 1 & 2 & 3 & 4 & 5 & 6 & 7 & 8 & 9 \\  \hline 
 \text{D2} & \times & \times & \times &  &  &  &  &  &  &  \\
N\, \text{D6}  & \times & \times & \times &  & \times & \times & \times &\times  &  &  
\end{array}
$
\end{center}

It is well known that the low-energy theory living on the probe is a 3d $\mathcal{N}=4$ field theory with $U(1)$ gauge group and $N$ hypermultiplets (originating from $2-6$ strings) transforming in the fundamental of the $U(N)$ flavor symmetry. We call $\{Q_i,\tilde{Q}^i\}_{i=1,\ldots,N}$ the chiral components of such hypermultiplets, with gauge charge $+1,-1$ respectively. Strings stretching from the D2 to itself, instead, give rise to a vector multiplet and a neutral hypermultiplet. The former comprises a vector field $A_\mu$, a complex scalar field $\phi$ describing the motion of the probe transverse to the  D6-stack, i.e. in the ($8,9$)-plane, and a real scalar field $\sigma$ for the motion along direction $3$. The latter, whose chiral halves we call $s_1,s_2$, is associated to the motion of the probe in the directions longitudinal to the D6-stack, namely along $4,5,6,7$, and in the $\cN=4$ theory is a free field. Interactions are described by a superpotential of the form
\be\label{3dSuperpot}
W=\sum_{i=1}^N\tilde{Q}^i\phi Q_i\,.
\ee

This theory has a non-trivial IR physics, and here we are mostly interested in the Coulomb branch of its moduli space, which can be described as follows \cite{Seiberg:1996bs,Intriligator:1996ex}. We first Hodge-dualize the photon to a real scalar $\gamma$: The latter is periodic and lives on a circle of radius equal to the square of the gauge coupling $g$. Then we cast the fields $\gamma,\sigma$ in the so-called ``monopole operators'':
\be
V_\pm\sim e^{\pm(\sigma+i\gamma/g^2)}\,.
\ee
The above are to be interpreted as classical relations, valid far out along the Coulomb branch, where they satisfy the obvious constraint $V_+V_-=1$. However, at distances of order $g^2$ from the origin, the Coulomb branch drastically deviates from a cylindrical shape, due to strong quantum corrections. For a definition of the monopole operators $V_{\pm}$ valid in the full quantum theory, see \cite{Borokhov:2002cg}. Quantum corrections turn the Coulomb branch into the following ALF space:
\be\label{TaubNut}
V_+V_-=\phi^N\,.
\ee  
This phenomenon is elegantly described by the M-theory lift of the above type IIA set-up, where the D6-branes precisely become the $N$-center Taub-NUT space described by \eqref{TaubNut} and probed by a M2-brane, while the gauge coupling gets ``geometrized'' into the radius of the 11th dimension. The IR fixed point of the theory on the probe corresponds to sending the gauge coupling to infinity, and thus to turning the Taub-NUT into $\mathbb{C}^2/\mathbb{Z}_N$, an ALE space with an $A_{N-1}$-type singularity at the origin. Given $\mathbb{C}^3$ with coordinates $u,v,z$, this space is conveniently modeled by the following holomorphic surface
\be\label{IRCB}
uv=z^N\,,
\ee
where $u,v$ are to be understood as ``fiber'' coordinates, whereas $z$ parametrizes the base, being identified with the field $\phi$. The M-theory configuration we are considering is summarized in the following table

\begin{center}
$
\begin{array}{c|ccccccccccc}
\text{M-theory}& 0 & 1 & 2 & 3 & 4 & 5 & 6 & 7 & 8 & 9 & 10\\  \hline 
 \text{M2} & \times & \times & \times &  &  &  &  &  &  &  \\
\text{ALE}  &  &  &  & \times &  &  &  &  & \times & \times  &\times
\end{array}
$
\end{center}

We are now interested in adding a specific class of field-dependent relevant deformations to the superpotential \eqref{3dSuperpot}, which can be described as
\be\label{Deformation}
\delta W= Tr(\mu M)\,,
\ee
where $\mu_i^j\equiv Q_i\tilde{Q}^j$ is the meson matrix (or the so-called ``moment map'' associated to the $U(N)$ flavor symmetry), and $M$ is a gauge-invariant chiral superfield that we are adding to the theory, transforming in the adjoint of the flavor group. The string-theory set-up we are using to engineer the field theory leads to interpret the extra field $M$ as the vacuum expectation value of the ``Higgs field'' $\Phi$ of the D6-branes, namely of the background field whose spectral data describe the motion of the D6-stack in the ($8,9$)-plane:
\be
M=\langle\Phi\rangle\,.
\ee
If $M$ is constant and such that $[M,M^\dagger]=0$, its entries are (complex) masses from the probe point of view, and the corresponding deformation \eqref{Deformation} preserves $\cN=4$ supersymmetry.
However, since $\langle\Phi\rangle$ can also depend on the internal coordinates of the D6-branes, which are identified with the singlets $s_1,s_2$ of the gauge theory, the deformation \eqref{Deformation} can also introduce new interactions, which only preserve four supercharges. Introducing the term \eqref{Deformation} in the superpotential triggers an RG-flow which will generally lift (part of) the Higgs branch and will deform the IR Coulomb branch as
\be\label{DefIRCB}
uv=P_M\,,
\ee
where $P_M\equiv \det(z\mathbb{1}-M)$ denotes the characteristic polynomial of the $N\times N$ matrix $M$.

In this paper, we will mostly be interested in the kind of deformations considered in 4d by Maruyoshi and Song \cite{Maruyoshi:2016tqk}, and thus will let $M$ acquire a nilpotent vacuum expectation value. As is well known, this breaks the adjoint representation of $SU(N)$ into a sum of $SU(2)$ representations, labeled by the spin. As a result, only the components $\delta M_{(j,-j)}$ will remain coupled, where the subscript denotes the lowest state of the representation with spin $j$. We will associate $s_1,s_2$ with the fluctuations $\delta M_{(j,-j)}$ with the two highest spins\footnote{As we will see especially in 4d, for some nilpotent orbits the maximum spin is multiply populated, in which case we will let $s_1,s_2$ be two fields with the same spin.}. It is immediate to see that these deformations originate from a Higgs-field background with the property that
\be
[\langle\Phi\rangle,\langle\Phi\rangle^\dagger]\neq0\,,
\ee
which correspond to having formed a T-brane of the D6-stack \cite{Cecotti:2010bp}. However, as we will see momentarily, these are the types of T-branes which do deform the geometry of the Coulomb branch: Indeed, due to the fluctuations, the spectral data of $M$ will not be empty.

Before starting to analyze the consequences of the RG flows generated by \eqref{Deformation}, it is worth remarking here that, clearly, this realization of the field theory forces $M$ to only contain at most two singlet fields. As we will see, for the purpose of discussing supersymmetry enhancement, this will be enough and will not constitute a limitation at all. Actually, in the 3d Abelian theory treated in this section, even retaining just one of the two singlets will mostly be enough to show when supersymmetry enhances, and in which cases, instead, the IR theory will inevitably posses only $\cN=2$ supersymmetry.

\subsection{SUSY enhancement}

Let us start from the simplest case, i.e. $N=2$. As we increase the number of flavors later, we are going to see an easy pattern allowing us to make general statements about the enhancement. The results we find are in agreement with the field-theory analysis of \cite{Benvenuti:2017lle,Benvenuti:2017kud}.

For $SU(2)$ there is only one non-trivial nilpotent orbit, and a single spin-$1$ field to be considered. The corresponding deformation matrix is
\be
M=\langle M\rangle+\delta M_{(1,-1)}=\left(\begin{array}{cc}0&m\\ 0 &0\end{array}\right)+\left(\begin{array}{cc}0&0\\ s &0\end{array}\right)\,,
\ee
where we made explicit the mass scale through the parameter $m$. Now, adding the deformation \eqref{Deformation} to the superpotential \eqref{3dSuperpot}, and integrating out the massive fields, we end up with the following effective superpotential
\be
W^{\rm eff} = \left(s-\frac{\phi^2}{m}\right)Q_1\tilde{Q}^2\,.
\ee
By flowing to the IR, we look at the system at decreasing energy scales, which effectively means sending to infinity the scale $m$. At the IR fixed point, therefore, the superpotential simply becomes
\be
W^{\rm IR}= s\, Q_1 \tilde{Q}^2\,,
\ee
which gives us back $\cN=4$ supersymmetry. Alternatively, one can also invoke the chiral ring stability criterion of \cite{Benvenuti:2017lle} to remove the second term in $W^{\rm eff}$: The F-term equation for $s$ sets to zero  $Q_1\tilde{Q}^2$ in the chiral ring and therefore the $\phi$-dependent term can be dropped without affecting the infrared dynamics.

\begin{figure}[h!]
	\begin{center}
		\begin{tikzpicture}
		\draw (0,0) parabola (3.5,3);
		\draw (0,0) parabola (-3.5,3);
		\draw [->, thick] (-4,0) -> (4,0);
		\draw [->, thick] (0,0) -> (0,4);
		\draw[red, thick] (-0.3,0.9) arc (160:200:2);
		\draw[red, thick] (0.3,-0.5) arc (-20:20:2);
		\draw [->, thick, red] (-1,-1.2) -> (0,-.2);
		\node at (4,-0.5) {$z$};
		\node at (0.5, 3.9) {$s$};
		\node[red] at (-1, -1.5) {IR window};		
		\end{tikzpicture}
		\caption{In the IR, the D2 only probes a tiny neighborhood around $s=z=0$ of the curve $s-z^2=0$, where the D6 appears to be flat.}
		\label{fig:Parabola}
	\end{center}
\end{figure}
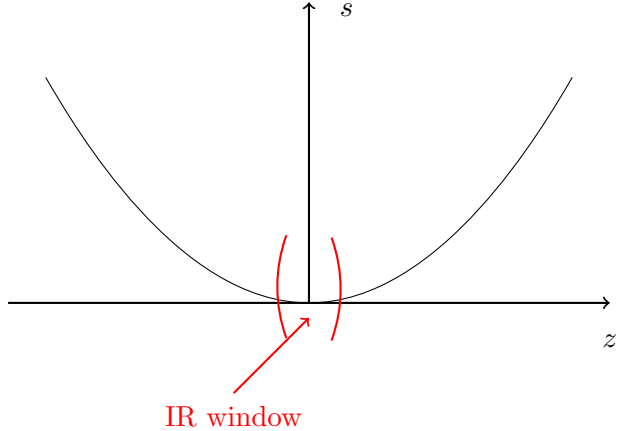

The final theory is however different from the starting one: Neglecting the decoupled chiral multiplet $\phi$, we have obtained a $\cN=4$ theory with a single fundamental hypermultiplet. This theory has trivial Higgs branch, and in the IR it is actually equivalent to the theory of a free hypermultiplet. Indeed, using \eqref{DefIRCB}, its IR Coulomb branch is
\be
uv = z^2-ms \qquad  \Longrightarrow  \qquad  uv = s\,,
\ee
where we have performed a trivial redefinition of the coordinate $s$. The latter can now be eliminated in favor of $u,v$, which correspond to the free hypermultiplet. Geometrically, this phenomenon of enhancement amounts to a rather obvious statement: The deformed Coulomb branch is a smooth threefold, and hence it can always be written locally as a twofold times a line. This has a clear physical meaning: The deformed theory is realized on a D2-brane probing a \emph{single} D6-brane wrapping the \emph{curved} space $z^2-s=0$, which has a parabolic shape in the complex plane with coordinates $z,s$ (see figure \ref{fig:Parabola}). In the IR, however, the D2 does not dispose of enough energy to ``feel'' the curvature of the D6, being only able to probe a tiny neighborhood of it around $s=0$. At the fixed point, the probe just sees a \emph{flat} D6-brane at $s=0$, which is the reason for supersymmetry enhancement.

The above discussion makes it clear that, whatever is the number of flavors we start with, if the $SU(N)$ flavor symmetry is completely broken by a $\langle M\rangle$ corresponding to the \emph{maximal} nilpotent orbit, the deformed $\cN=2$ theory always flows to a free $\cN=4$ one. The argument proceeds through the same steps: Under the principal embedding, the adjoint representation of $SU(N)$ breaks as follows \cite{Gadde:2013fma,Agarwal:2014rua}
\be
{\rm Adj}_{SU(N)} \quad\longrightarrow\quad \bigoplus_{j=1}^{N-1} \,V_j\,,
\ee
where $V_j$ denotes the $SU(2)$ representation of spin $j$. Now, in principle, we should retain all of the $N-1$ spin $(j,-j)$ fields, which will give rise to several interaction terms in the deformed superpotential. It is immediate to see, however, that the dominant term in the $m\to\infty$ limit is always the one involving the highest spin, which, as before, substitutes $\phi$ in the role of complex scalar in the $\cN=4$ vector multiplet. The physical meaning and the geometric realization of the enhancement are identical to the case discussed above\footnote{The differences are just in higher-order monomials near the origin, neglected at the IR fixed point. For instance, retaining only the two highest spins, the Coulomb branch of the deformed theory is a fourfold of the form $uv=z^N + m^{N-2} s_2 z + m^{N-1} s_1$, which locally splits in $\{uv=s_1\}\times\mathbb{C}^2_{z,s_2}$.}.

For any nilpotent orbit other than the maximal one, the RG flows triggered by deformations analogous to those considered by Maruyoshi and Song in 4d do not lead to supersymmetry enhancement. Let us give two slightly different examples of non-principal embeddings, from which a general pattern can be easily inferred. First consider $N=3$ in the starting theory, and deform it according to the subregular (or minimal) nilpotent orbit of $SU(3)$, i.e. with
\be
M=\left(\begin{array}{ccc}0&m&0\\ s &0&0\\ 0&0&0\end{array}\right)\,,
\ee
where again we only kept the field of highest spin, $(1,-1)$. The Coulomb branch of the deformed geometry has now a singularity of conifold type:
\be
uv=z^3-msz \qquad  \Longrightarrow  \qquad  uv = sz\,,
\ee
and consequently the M2 probing this geometry only preserves $\cN=2$ supersymmetry. Analogously, in type IIA, the D2 sits near the intersection point of two D6-branes.

As a second example, take $SU(4)$ with the deformation induced by
\be\label{defsu4}
M=\left(\begin{array}{cccc}0&m&0&0\\ s_1 &0&0&0\\ 0&0&0&m\\ 0&0&s_2&0 \end{array}\right)\,,
\ee
which descends from the $[2,2]$ partition of $4$. Here, in the decomposition of the adjoint, one finds four spin-$1$ fields, and hence we need to retain at least a couple of them, $s_1$ and $s_2$. Again, at low energies, the probe sees a conifold geometry:
\be
uv=(z^2-ms_1)(z^2-ms_2)\qquad  \Longrightarrow  \qquad  uv = s_1s_2\,,
\ee
and the IR fixed point only has four conserved supercharges. Interestingly, notice that if we had considered a slightly different deformation of this theory, namely with $M$ like in \eqref{defsu4}, but with $s_1$ and $s_2$ identified, we would have found supersymmetry enhancement in the IR, with a non-trivial fixed point: a M2-brane probing the singular space $\mathbb{C}^2/\mathbb{Z}_2$.

It is clear that, in general, for all non-principal embeddings, we end up having at least two chiral fields which remain coupled in the IR theory. Consequently there is no supersymmetry enhancement along the RG flow.

\section{Rank-1 four-dimensional theories}
\label{sec:4d}

In this section we turn our attention to 4d $\cN=2$ supersymmetric field theories of rank $1$. The reason why we focus on them is that they have an easy geometric realization in the context of F-theory, which will help us explore the question of supersymmetry enhancement in purely algebraic terms. Our results agree and provide an explanation of the phenomena observed in \cite{Maruyoshi:2016aim,Agarwal:2016pjo,Giacomelli:2017ckh} using field-theoretic methods. Moreover, for the RG flows leading to supersymmetry enhancement, we are able to compute algebraically the correct IR conformal dimensions of all fields, i.e. without using any maximization procedure.

\subsection{Geometric set-up}

The geometric engineering of 4d $\cN=2$ field theories and of their $\cN=1$ deformations in F-theory is closely related to the M-theory setting we used to engineer 3d theories in Section \ref{ss:geo3d}. There are, however, a few important differences which we would like to highlight here.

Rank-$1$ field theories can be realized as theories on a single D3-brane probing a stack of 7-branes in type IIB string theory \cite{Banks:1996nj}. Contrary to the previous case, the 7-branes can be mutually non-perturbative, thus realizing exceptional flavor symmetries\footnote{or giving a non-perturbative realization of classical flavor symmetries, as is the case for the Argyres-Douglas theories.}. The corresponding probe theories are the so-called Minahan-Nemeschansky theories \cite{Minahan:1996cj}. F-theory provides the set-up to analyze these systems with geometric methods. It uses an auxiliary 2-torus fibered over the physical space, which from the field-theory viewpoint plays the role of the Seiberg-Witten curve. As in M-theory, one probes isolated singularities of elliptically-fibered ALE spaces, which will now be of different Kodaira types, according to the flavor structure. Probe and singular space for the starting theory in the UV extend in twelve dimensions as follows:

\begin{center}
$
\begin{array}{c|cccccccccccc}
\text{F-theory}& 0 & 1 & 2 & 3 & 4 & 5 & 6 & 7 & 8 & 9 & 10 &11\\  \hline 
 \text{D3} & \times & \times & \times &\times  &  &  &  &  &  &  \\
\text{ALE}  &  &  &  &  &  &  &  &  & \times & \times  &\times &\times
\end{array}
$
\end{center}
In the above table, $10$ and $11$ represent the torus directions and, unlike in M-theory, do not correspond to any physical operator of the probe theory. Instead, the ($8,9$)-plane, which the torus is fibered over, is parametrized by the UV Coulomb-branch operator (this is the motion of the D3 probe transverse to the 7-branes). Like in the previous section, the $4,5,6,7$ coordinates, $s_1,s_2$, correspond to a free hypermultiplet parameterizing the motion of the probe along the 7-brane stack, which will be coupled when deforming the theory.

Deformations are again formulated as in \eqref{Deformation}, where $\mu$ is the moment map associated to the flavor symmetry of the starting theory, which in most of the cases is a non-Lagrangian theory. $M$ is the extra chiral field added to the theory, which, following \cite{Gadde:2013fma}, we will split as
\be\label{MSmatrix}
M=\rho(\sigma_+) + \sum_{j=s_1,s_2} \delta M_{(j,-j)}\,,
\ee
where the first term is its vacuum expectation value, taken along the nilpotent element $\rho(\sigma_+)$. Again, for the purpose of discussing the possible appearance of supersymmetry enhancement in the IR, it will be sufficient to restrict the above sum of fluctuations to the two highest spins, $s_1, s_2$, in the decomposition of the adjoint representation of the original flavor symmetry. We name them such that spin$(s_1)\geq$ spin$(s_2)$. The UV conformal dimension $D^{\rm UV}(\cdot)$ of such fields is related to their spin by \cite{Gadde:2013fma} (see also \cite{Maruyoshi:2016aim})
\be\label{DvsSpin}
D^{\rm UV}(s_i)={\rm spin}(s_i) +1\,,
\ee
as we will review in Appendix \ref{Scan}. Hence fields of vanishing spin are free fields.

Like in the M-theory construction, $M$ induces a deformation of the ALE space in F-theory, because it corresponds to the vacuum expectation value of the 7-brane Higgs-field. The probed configuration is again of T-brane type. In this case, however, there is a technical complication: The characteristic polynomial of $M$ does not directly appear in the one defining the geometry, like in \eqref{DefIRCB}. Nevertheless, there exists a precise one-to-one correspondence between the Casimir invariants of $M$ and the versal\footnote{Roughly said, these are all deformations which cannot be undone by coordinate redefinitions.} deformations of the original singular geometry (see e.g. \cite{Katz92}). See Table \ref{table1}, taken from \cite{Argyres:2015gha}, for a summary of the complete unfolding of the singularities which are relevant to us in this section. By using this correspondence, we are able to write down the deformed F-theory geometry for any given nilpotent orbit.

\begin{table}[h!]
\centering
\scalebox{0.9}{
\begin{tabular}{|c|c|c|}
\hline 
Kodaira & Surface & Flavor  \\ 
\hline 
$II^{*}$ & $y^2=x^3+x(M_2z^3+M_8z^2+M_{14}z+M_{20})+(z^5+M_{12}z^3+M_{18}z^2+M_{24}z+M_{30})$  & $E_8$ \\ 
$III^{*}$ & $y^2=x^3+x(z^3+M_{8}z+M_{12})+(M_2z^4+M_6z^3+M_{10}z^2+M_{14}z+M_{18})$  & $E_7$\\ 
$IV^{*}$ & $y^2=x^3+x(M_2z^2+M_5z+M_8)+(z^4+M_6z^2+M_9z+M_{12})$ &  $E_6$\\ 
$I_0^{*}$ & $y^2=x^3+x(\tau z^2+M_2z+M_4)+(z^3+\tilde{M}_4z+M_6)$  & $SO(8)$\\ 
$IV$ & $y^2=x^3+xM_2+(z^2+M_3)$  & $SU(3)$\\ 
$III$ & $y^2=x^3+xz+M_2$  & $SU(2)$\\ 
$II$ & $y^2=x^3+z$  & no\\ 
\hline 
\end{tabular} }
\caption{Maximally deformed singularities. $M_i$ is the degree-$i$ Casimir invariant of the corresponding flavor symmetry. $\tilde{M}_4$ indicates the Pfaffian of the $\mathfrak{so}(8)$ matrix. Table taken from \cite{Argyres:2015gha}.}
\label{table1}
\end{table}

Regardless of which RG flow and which energy scale we look at, the probed F-theory geometry is always going to look like a hypersurface in Weierstrass form
\be\label{genWeier}
y^2=x^3+f(z,s_1,s_2)x+g(z,s_1,s_2)\,,
\ee
where $x,y$ are the auxiliary fiber directions and $z$ parametrizes the UV Coulomb branch. The precise form of the holomorphic functions $f,g$ depends on the choice of UV theory we start with, and on the way we deform it (i.e. on the choice of nilpotent orbit in \eqref{MSmatrix}). The ALE space corresponding to the UV theory is retrieved from \eqref{genWeier} by setting $s_1\equiv s_2\equiv 0$.

\subsection{SUSY enhancement}

Let us now perform an algebro-geometric analysis of the RG flows triggered by the T-brane deformations described above. In particular, our method will tell us which nilpotent orbits are expected to lead to supersymmetry enhancement in the IR, and what are the $\cN=2$ geometries we are supposed to land on in each case. Results are in agreement with \cite{Maruyoshi:2016aim,Agarwal:2016pjo,Giacomelli:2017ckh}.

\subsubsection{Approach}

The logic is the same of the previous section: If supersymmetry enhances in the IR, the deformed geometry, given by the (local) Calabi-Yau fourfold in equation \eqref{genWeier}, must factorize into a Calabi-Yau twofold and a trivial factor:
\be
{\rm CY^{\rm UV}_2} \qquad \xRightarrow[\phantom{XXX}]{{\rm Def}} \qquad{\rm CY_4} \qquad \xRightarrow[\phantom{XXX}]{{\rm IR}} \qquad {\rm CY^{\rm IR}_2}\; \times \;\mathbb{C}^2\,,
\ee
Here we made it manifest that the twofold geometry we find in the IR is generally different from the one we start with in the UV. The two are obviously the same if we choose the trivial nilpotent orbit, whereby $s_1,s_2$ are both free fields (being of spin $0$), and thus $\delta W\equiv0$. In fact, as we will see momentarily, this is the only instance for which the twofold geometries coincide.

To proceed, we employ the following strategy: First, since conformal dimensions of operators may be viewed as $\mathbb{C}^*$-assignments for the corresponding algebraic variables, we promote the affine coordinates in \eqref{genWeier} to projective ones, and require the fourfold polynomial to be homogeneous. Homogeneity is indeed the geometric counterpart of the fact that relative scalings of the fields in question are invariant under RG flow\footnote{See Appendix \ref{Scan} for a field-theory proof of this  fact.}. 

Next, we assume that supersymmetry enhances at the end of the flow and we make an ``educated guess'' of what the IR twofold geometry would be. Our Ansatz is
\be\label{Ansatz}
{\rm CY^{\rm IR}_2} = {\rm CY_4}|_{z\equiv s_2\equiv0}\,,
\ee
that means promoting the highest-spin field $s_1$ to the role of IR Coulomb-branch operator (or equivalently making it the base coordinate for the elliptic ${\rm CY^{\rm IR}_2}$). Note that, on the one hand, choosing to retain $s_2$ instead of $s_1$ would have not been a viable option because, due to homogeneity, a decoupling of $s_1$ would force $s_2$ to decouple too. On the other hand, suppressing $s_1,s_2$, namely imposing ${\rm CY^{\rm IR}_2}={\rm CY^{\rm UV}_2}$, would force $z$ to maintain its UV conformal dimension, again due to homogeneity. This in turn implies trivial RG flow, and thus trivial nilpotent orbit.

Finally, we must make sure that our Ansatz \eqref{Ansatz} is indeed consistent. In order to do so, we compute the IR conformal dimensions of the various fields and verify that both $z$ and $s_2$ will hit the unitarity bound and decouple \cite{Kutasov:2003iy}, which means:
\be\label{unitarity}
D^{\rm IR}(z)\leq1  \qquad {\rm and}\qquad  D^{\rm IR}(s_2)\leq1\,.
\ee
We argue that the deformations for which this happens lead to IR supersymmetry enhancement. In contrast, when at least one of these fields does not satisfy \eqref{unitarity}, the IR fixed point certainly preserves only four supercharges. 

Based on the above considerations, we can immediately conclude that a non-trivial orbit characterized by an adjoint decomposition where the highest-spin state is multiply populated, can never lead to supersymmetry enhancement. This is because $D(s_1)=D(s_2)$ both in UV and IR, and thus the second relation in \eqref{unitarity} cannot be satisfied. Similar conclusion holds for those orbits where the highest-spin field has UV dimension smaller or equal to that of the Coulomb-branch operator, because the first relation in \eqref{unitarity} would be violated.

Conformal dimensions in the IR are computed in a purely algebraic manner, by exploiting the assumption of extended supersymmetry:\footnote{We can of course compute also the UV conformal dimensions by the same method.} As is usually done to construct local models in F-theory \cite{Beasley:2008dc}, we imagine the twofold ${\rm CY^{\rm IR}_2}$ fibered over the rest of the space, and thus let the homogeneous coordinates $x,y,s_1$ to be sections of suitable powers of the canonical bundle of the base. These powers are the would-be conformal dimensions in the IR. We now impose that the total space of such fibration be Calabi-Yau, which, using adjunction, amounts to the condition\footnote{As explained in Appendix \ref{Scan}, this condition is a geometric counterpart of the field-theory fact that the Seiberg-Witten differential has dimension $1$.}
\be\label{CYcond}
D^{\rm IR}(x)-D^{\rm IR}(y)+D^{\rm IR}(s_1)=1\,.
\ee
Together with homogeneity, this equation allows us to determine $D^{\rm IR}$ for all fields. They will coincide with the correct IR conformal dimensions, in case \eqref{unitarity} is satisfied and thus the assumption of enhancement is found to be consistent. In the cases where we find no enhancement, instead, this algebraic method is unreliable, and one cannot bypass the a-maximization process to derive the correct IR scaling of fields.

We will now exemplify the strategy outlined above in three qualitatively distinct cases of enhancing RG flows and in one instance without enhancement, deferring to Appendix \ref{Scan} a complete scan of rank-$1$ theories and of their Maruyoshi-Song deformations.

\subsubsection{SU(2) gauge theory with 4 flavors and maximal orbit}

Among the UV theories we consider, this is the only case with a weakly-coupled Lagrangian description, and the analysis can be performed entirely in a type IIB setting of a D3-brane probing a stack of 4 D7-branes attached to an O7$^-$ plane. We will however proceed using the more powerful geometric setting of F-theory, which can be exported to the strongly coupled cases of the next subsections.

The $SU(2)$ gauge theory with $SO(8)$ flavor symmetry arises as the theory on a D3-brane probing an elliptically-fibered ALE space with D$_4$ singularity at the origin (Kodaira type I$^*_0$). The corresponding hypersurface equation in $\mathbb{C}^3$ reads
\be\label{D4}
y^2 = x^3 + \tau z^2 x + z^3\,,
\ee
where $\tau$ denotes the exactly marginal coupling of this theory. As discussed, we can derive the UV conformal dimension of the Coulomb-branch operator $z$ and the scalings of the auxiliary variables $x,y$, by simply fibering this singular space over the 7-brane worldvolume. This gives us the following three conditions:
\begin{eqnarray}
D^{\rm UV}(x) - D^{\rm UV}(y) + D^{\rm UV}(z) &=& 1\,,\\
3D^{\rm UV}(x)- 2 D^{\rm UV}(y)&= &0\,, \\
D^{\rm UV}(x)-D^{\rm UV}(z)&=&0\,,
\end{eqnarray}
where the first comes from the Calabi-Yau condition of the total space of the fibration (the UV analog of \eqref{CYcond}), whereas the others are simply consequence of the homogeneity of \eqref{D4}. Solving the above system gives $D^{\rm UV}(z)=D^{\rm UV}(x)=2$ and $D^{\rm UV}(y)=3$.

We are now interested in deforming this theory with $\langle M\rangle$ in the principal nilpotent orbit of $\mathfrak{so}(8)$. The decomposition of the adjoint corresponding to this orbit reads (see Table \ref{D4orbits}):
\be
{\rm Adj} \longrightarrow V_5 \oplus V_3 \oplus V_3 \oplus V_1\,,
\ee
where $V_j$ indicates the representation of spin $j$ under the embedded $SU(2)$. We now retain only two of the four extra singlets which remain coupled to the theory, namely the spin-$5$ field and one of the spin-$3$ fields, and identify them with $s_1,s_2$ respectively. The deformation we consider is then of the form \eqref{Deformation}, with $M$ the following $8\times8$ matrix
\be\label{MforSO8}
M=\left(
\begin{array}{cccccccc}
0 & \sqrt{6} & 0 & 0 & 0 & 0 & 0 & 0 \\
0 & 0 & \sqrt{10} & 0 & 0 & 0 & 0 & 0 \\
0 & 0 & 0 & \sqrt{6} & 0 & 0 & 0 & \sqrt{6} \\
s_2 & 0 & 0 & 0 & 0 & 0 & -\sqrt{6} & 0 \\
0 & s_1 & 0 & s_2 & 0 & 0 & 0 & -s_2 \\
-s_1 & 0 & 0 & 0 & -\sqrt{6} & 0 & 0 & 0 \\
0 & 0 & 0 & 0 & 0 & -\sqrt{10} & 0 & 0 \\
-s_2 & 0 & 0 & 0 & 0 & 0 & -\sqrt{6} & 0 \\
\end{array}
\right)
\ee
For a discussion on how to build explicit standard triples for nilpotent orbits of complex simple Lie algebras, see for example \cite{McGovern:2296111}.
The characteristic polynomial of the above matrix reads
\be
P_M(t)=t^8+240\sqrt{6} s_1 t^2+ (120 s_2)^2\,,
\ee
from which we see that two of the four independent Casimir invariants of $\mathfrak{so}(8)$ are activated, i.e. the sixth-order Casimir and the Pfaffian (one of the two fourth-order Casimir's). After a suitable rescaling, they can be identified with $s_1,s_2$ respectively. We can now derive the versal deformations of the D$_4$ singularity induced by \eqref{MforSO8}, finding the (smooth) fourfold (see Table \ref{table1}):
\be\label{D4-H0}
y^2=x^3+\tau z^2x+z^3+s_2z+s_1\,.
\ee
Following our procedure, we make at this point the assumption that this RG flow leads to supersymmetry enhancement and make the Ansatz \eqref{Ansatz} for the twofold in the IR, i.e.
\be\label{H0}
{\rm CY^{\rm IR}_2}\,: \qquad y^2=x^3+s_1\,.
\ee
This is a smooth local elliptic K3 manifold, with a singular fiber in the origin of cusp form (Kodaira type II). We now compute the new scaling dimensions of $x,y,s_1$, by using \eqref{CYcond} and homogeneity of \eqref{H0}, and we find $D^{\rm IR}(x)=2/5$, $D^{\rm IR}(y)=3/5$, $D^{\rm IR}(s_1)=6/5$. Homogeneity of \eqref{D4-H0} then fixes the conformal dimensions of $z,s_2$ to be $D^{\rm IR}(z)=2/5$, $D^{\rm IR}(s_2)=3/5$, which violate unitarity unless the fields $z,s_2$ decouple. This makes our Ansatz consistent, and we have found that the $\cN=2$ IR fixed point is the theory of a D3-brane probing the space \eqref{H0}, i.e. the Argyres-Douglas theory of type $\mathcal{H}_0$. Also, the dimension of its Coulomb-branch operator $s_1$ ($6/5$) turns out to be the correct one.

\subsubsection{${\bf E_7}$ Minahan-Nemeschansky and ${\bf E_6}$-type orbit}

The starting UV theory here is the one living on a D3-brane probing the space
\be
y^2 = x^3 + xz^3\,,
\ee
i.e. a local K3 surface with an E$_7$ singularity in the origin (Kodaira type III$^*$). Following the usual trick, we compute the scaling dimensions of $x,y,z$, finding $D^{\rm UV}(x)=6$, $D^{\rm UV}(y)=9$ and $D^{\rm UV}(z)=4$. The deformation we would like to consider corresponds to the nilpotent orbit of $\mathfrak{e}_7$ with Bala-Carter label E$_6$, which has an adjoint decomposition such that (see Table \ref{tableE7})
\be
{\rm Adj} \;\supset \; V_{11} \oplus V_8 \,.
\ee
We therefore identify the spin-$11$ and the spin-$8$ fields with $s_1,s_2$ respectively. Among the $7$ independent Casimir invariants of the $\mathfrak{e}_7$ algebra, there is one of degree $12$ and one of degree $18$. By a quick scaling argument which uses equation \eqref{DvsSpin}, we conclude that both of them are activated and are proportional to $s_1$ and $s_2^2$ respectively. There are no other combinations of the two singlets with a degree matching that of any other Casimir invariants. As can be seen from Table \ref{table1}, the deformed hypersurface is then (modulo rescaling of fields)
\be\label{E7-H1}
y^2 = x^3 + xz^3 +s_1x+s_2^2\,.
\ee
We now conjecture that supersymmetry enhances with a IR twofold of the form:
\be\label{H1}
{\rm CY^{\rm IR}_2}\,: \qquad y^2=x^3+s_1x\,,
\ee
which gives us the following IR conformal dimensions: $D^{\rm IR}(x)=2/3$, $D^{\rm IR}(y)=1$, $D^{\rm IR}(s_1)=4/3$. Homogeneity of \eqref{E7-H1} fixes $D^{\rm IR}(z)=4/9$ and $D^{\rm IR}(s_2)=1$, implying that both $z$ and $s_2$ decouple. Our conjecture is indeed verified and we have found in the IR the Argyres-Douglas theory of type $\mathcal{H}_1$ with the correct conformal dimension for its Coulomb-branch operator $s_1$. This theory arises on a D3-brane probing the elliptic surface \eqref{H1} with singularity of Kodaira type $III$.

\subsubsection{${\bf E_6}$ Minahan-Nemeschansky and ${\bf D_4}$-type orbit}
Here we start from the theory of a D3-brane probing the following ALE space
\be
y^2 = x^3 + z^4\,,
\ee
which has an E$_6$ singularity in the origin (Kodaira type IV$^*$). Scaling dimensions in the UV are: $D^{\rm UV}(x)=4$, $D^{\rm UV}(y)=6$ and $D^{\rm UV}(z)=3$. We consider the deformation associated to the nilpotent orbit of $\mathfrak{e}_6$ with Bala-Carter label D$_4$, which has an adjoint decomposition with the property (see Table \ref{tableE6})
\be
{\rm Adj} \;\supset \; V_{5} \oplus V_3 \,.
\ee
We are led to identify the spin-$5$ field with $s_1$ and the spin-$3$ field with $s_2$. Among the $6$ independent Casimir invariants of the $\mathfrak{e}_6$ algebra, there is one of degree $8$ and one of degree $12$, which we identify with $s_2^2$ and $s_1^2$ respectively. By a degree-argument we can conclude that no other Casimir is activated. Using Table \ref{table1}, we can write the deformed geometry as
\be\label{E6-H2}
y^2 = x^3 + z^4 +s_2^2x+s_1^2\,.
\ee
Our educated guess for the IR twofold corresponding to the susy-enhanced theory is
\be\label{H2}
{\rm CY^{\rm IR}_2}\,: \qquad y^2 = x^3+s_1^2\,,
\ee
which leads to the IR dimensions: $D^{\rm IR}(x)=1$, $D^{\rm IR}(y)=3/2$, $D^{\rm IR}(s_1)=3/2$. Our guess is justified by the fact that the former Coulomb-branch operator $z$ and the extra singlet $s_2$ hit the unitarity bound and decouple. Indeed, using homogeneity of \eqref{E6-H2}, we have $D^{\rm IR}(z)=3/4$ and $D^{\rm IR}(s_2)=1$. We have found in the IR the Argyres-Douglas theory of type $\mathcal{H}_2$, as the theory on a D3-brane probing the singular ALE space \eqref{H2} with type-IV Kodaira singularity.

\subsubsection{${\bf E_8}$ Minahan-Nemeschansky and ${\bf E_8(a_2)}$-type orbit}

We would like to conclude this section by giving an example for which our procedure guarantees the absence of IR supersymmetry enhancement. We start in the UV with the E$_8$ Minahan-Nemeschansky theory, i.e. with a D3-brane probing the ALE space
\be\label{E8}
y^2 = x^3 + z^5\,,
\ee
a singular space with Kodaira-type singularity II$^*$. Computing the UV scaling dimensions, we find: $D^{\rm UV}(x)=10$, $D^{\rm UV}(y)=15$ and $D^{\rm UV}(z)=6$. Consider the deformation induced by the nilpotent orbit of $\mathfrak{e}_8$ with Bala-Carter label E$_8$(a$_2$), which admits an adjoint decomposition such that (see Table \ref{tableE8p1})
\be
{\rm Adj} \;\supset \; V_{19} \oplus V_{17} \,.
\ee
We identify the spin-$19$ and the spin-$17$ fields with $s_1,s_2$ respectively. Among the $8$ independent Casimir invariants of $\mathfrak{e}_8$, the only ones which are turned on by this deformations are the one of degree $18$ and the one of degree $20$, which, after suitable field redefinitions, can be taken to coincide with $s_2$ and $s_1$ respectively. The geometry \eqref{E8} will then be deformed as follows (see Table \ref{table1})
\be\label{E8-/}
y^2 = x^3 + z^5 + s_1 x + s_2 z^2\,.
\ee
We now assume that this RG flow leads to supersymmetry enhancement and make the following Ansatz for the IR twofold:
\be
{\rm CY^{\rm IR}_2}\,: \qquad  y^2 = x^3 + s_1 x \,,
\ee
which is an ALE space with type-III Kodaira singularity, just like \eqref{H1}. Hence, IR scaling dimensions are: $D^{\rm IR}(x)=2/3$, $D^{\rm IR}(y)=1$, $D^{\rm IR}(s_1)=4/3$. Using homogeneity of \eqref{E8-/}, however, we find that the new scaling dimensions of $z,s_2$ are: $D^{\rm IR}(z)=2/5$ and $D^{\rm IR}(s_2)=6/5$. Therefore, while the former Coulomb-branch decouples, the other extra singlet remains coupled and thus invalidates our assumption of IR supersymmetry enhancement. We can then definitely conclude that the IR theory has only $\cN=1$ supersymmetry, but the conformal dimensions we have computed are incorrect. In this case, we cannot bypass a differential process to derive them.

\section{Conclusions}
\label{sec:conclu}

In this note we have explained how the geometric realization of theories with $8$ supercharges in M/F-theory can be used to understand the phenomenon of supersymmetry enhancement upon a superpotential deformation of the Maruyoshi-Song type. The key fact is that, in the stringy setup we consider, the superpotential deformation is encoded in the Weierstrass polynomial (\ref{genWeier}), and therefore we gain control over the entire RG flow using purely geometric techniques. 

In our setup the phenomenon of supersymmetry enhancement is translated into the simple geometric constraint that the background should preserve half of the supersymmetry, and thus reduce at low energy to a twofold times a flat euclidean space. This requirement precisely provides the consistency conditions which select the class of T-branes inducing supersymmetry enhancement. 

Looking beyond the scope of this paper, our analysis potentially leads to some interesting observations regarding nilpotent T-branes. We make an example to explain this point: Take the $\mathcal{H}_1$ theory which flows to $\mathcal{H}_0$ upon a nilpotent vev for the $SU(2)$ adjoint. If we flip the singlet $s_1$ (thus making it massive),\footnote{By ``flipping an operator $\mathcal{O}$'' we mean adding by hand a chiral multiplet $s$ and turning on the superpotential term $\mathcal{W}=s\mathcal{O}$ \cite{Dimofte:2011ju}.} we are left at low energy with the same theory we would get by activating a constant nilpotent T-brane. The relation with the $\mathcal{H}_0$ theory clearly tells us that the IR fixed point can also be reached starting from a different background (the Kodaira-type II singularity realizing the $\mathcal{H}_0$ theory) and deforming the theory by flipping the Coulomb-branch operator. A relation of this type extends to several other cases of nilpotent T-branes: The low-energy theory living on the probe can be realized starting from a different geometry via a superpotential deformation. It would be interesting to investigate this aspect further and understand how general this phenomenon is. We hope to come back to this issue in the future.

Also, part of our construction extends to higher-rank theories which can be engineered in M-theory by wrapping M5-branes on a Riemann surface: The models discussed in this paper are just special cases. It would be important to extend our construction to that class of theories as well, and to formulate a precise geometric criterion for supersymmetry enhancement. A similar question can be asked in the more general context of class $\mathcal{S}$ theories. We are currently investigating these topics.

\bigskip

\centerline{\bf \large Acknowledgments}

We would like to thank A. Collinucci, H. Hayashi and D. Morrison for useful discussions. We would like to acknowledge the Galileo Galilei Institute for Theoretical Physics for hospitality during the initial stage of this project. RS is grateful to the Aspen Center for Physics (which is supported by National Science Foundation grant PHY-1607611), for hospitality during the final stage of this work. The work of FC is supported through a fellowship of the international programme ``La Caixa-Severo Ochoa", and the grants FPA2015-65480-P (MINECO/FEDER EU) of the “Centro de Excelencia Severo Ochoa” Programme, and the ERC Advanced Grant SPLE under contract ERC-2012-ADG-20120216-320421. The work of SG is partly supported by the INFN Research Project ST$\&$FI. The work of RS is supported by the program ``Rita Levi Montalcini'' for young  researchers (D.M. n. 975, 29/12/2014).

\bigskip

\newpage

\appendix

\section{Scan of rank-$\bf{1}$ theories}\label{Scan}

In this Appendix we formulate a criterion for supersymmetry enhancement using arguments based on the Seiberg-Witten (SW) curve, and then show explicitly that our criterion selects precisely for each theory the correct nilpotent orbits. We do not discuss in detail the RG flows starting from the Argyres-Douglas theories $\mathcal{H}_1$ and $\mathcal{H}_2$, which have already been discussed several times in the literature. It can be easily checked that in these two cases all choices of nilpotent vev pass our test and lead to supersymmetry enhancement in the IR. 

\subsection{RG flow invariance of relative scalings and enhancement criterion} 

As we have explained in Section \ref{sec:4d}, the SW curve for the models we consider in the present paper is an elliptic curve that can be uniformly written as in (\ref{genWeier}): 
\be\label{curvasw}
y^2=x^3+xf(z,s_1,s_2)+g(z,s_1,s_2)=0\,,
\ee 
where the precise form of $f$ and $g$ depend on the theory and the choice of nilpotent orbit. The curve associated with the UV theory, before turning on the mass deformation, is obtained simply setting $s_1=s_2=0$ in (\ref{curvasw}), and in this case $z$ describes the expectation value of the Coulomb-branch (CB) operator. The explicit form of the SW differential $\lambda_{\rm SW}$ is also model dependent. However, its derivative w.r.t. the CB operator is always equal to the unique (up to exact terms) holomorphic differential $dx/y$ of the torus \cite{Seiberg:1994rs}. This condition follows from extended supersymmetry. We therefore have (in the UV) the relation 
\be\label{scaleuv}
\frac{\partial\lambda_{\rm SW}}{\partial z}=\frac{dx}{y}\,.
\ee
As is well known, from the curve and differential one can extract the scaling dimensions of CB operators \cite{Argyres:1995xn}: By requiring homogeneity of the curve we can fix the dimension at the CFT point of all the variables up to a rescaling. In the case at hand, we conclude from (\ref{curvasw}) that 
\be\label{scale2}
2D(y)=3D(x)\,;\quad D(f)=2D(x)\,;\quad D(g)=3D(x)\,,
\ee 
and knowing the explicit form of $f$ and $g$ we can case-by-case write the dimensions of $s_1$, $s_2$ and $z$ in terms of a single unknown, say $D(x)$.
The overall scaling can then be determined, as already mentioned in Section \ref{sec:4d}, by exploiting the fact that the SW differential has dimension one. Equivalently, from (\ref{scaleuv}) we recover (\ref{CYcond}) 
\be\label{diffuv} 
1-D(z)^{\rm UV}=D(x)^{\rm UV}-D(y)^{\rm UV}=-\frac{D(x)^{\rm UV}}{2}\,,
\ee 
where in the second equality we have used (\ref{scale2}).  

The key observation for us is that the \emph{relative} scaling dimensions do not vary along the flow and are therefore the same in the UV and in the IR. This can be seen for example by noticing that the curve (\ref{curvasw}) describes both the UV and the IR theory and the curve is the only information we need to extract relative scaling dimensions. We can also provide a more direct field-theoretic argument based on symmetries, as we will now explain. The deformation (\ref{MSmatrix}) leaves two $U(1)$ symmetries unbroken: The Cartan of $SU(2)_R$ (which we denote $I_3$) and $r-2\rho(\sigma_3)$, where $r$ is the generator of $U(1)_{R}$ and $\rho(\sigma_3)$ is the Cartan of the $SU(2)$ subgroup of the flavor symmetry associated with the given nilpotent orbit. The infrared R-symmetry can be parametrized as follows (see \cite{Maruyoshi:2016aim} for conventions): 
\be\label{trrr} 
\mathcal{R}_{\epsilon}=(1+\epsilon)I_3+\frac{1-\epsilon}{2}\left(r-2\rho(\sigma_3)\right)\,,
\ee
where the value of $\epsilon$ can be determined via a-maximization. All the operators appearing in (\ref{curvasw}) are uncharged under $I_3$ and their trial R-charge is of the form $R_{\epsilon}(\mathcal{O})=(1-\epsilon)D^{\rm UV}(\mathcal{O})$. The case of singlets deserves some further comments: Their charge under $r$ is 2 and the charge under $\rho(\sigma_3)$ is equal to the spin of the corresponding $SU(2)$ representation. Therefore, we recover (\ref{DvsSpin}): $D^{\rm UV}(s_i)={\rm spin}(s_i) +1$. Clearly, the actual R-charge in the IR, and hence the dimension, depends on the value of $\epsilon$. However, the ratio of scaling dimensions does not depend on this quantity and coincides with the UV value. This argument clearly applies to the auxiliary coordinates $x$ and $y$ as well. In this sense, relative dimensions are RG invariant, and in what follows we will exploit the following relations: 
\be\label{uvir}
\frac{D^{\rm IR}(z)}{D^{\rm IR}(x)}=\frac{D^{\rm UV}(z)}{D^{\rm UV}(x)}\,;\qquad \frac{D^{\rm IR}(s_1)}{D^{\rm IR}(x)}=\frac{D^{\rm UV}(s_1)}{D^{\rm UV}(x)}\,;\qquad \frac{D^{\rm IR}(s_2)}{D^{\rm IR}(x)}=\frac{D^{\rm UV}(s_2)}{D^{\rm UV}(x)}\,.
\ee 
In the above equation, by $D^{\rm IR}(\cdot)$ we actually mean the charge under $r/2-\rho(\sigma_3)$, which coincides with the actual scaling dimension in the IR only for operators which do not hit the unitarity bound and decouple. Our goal is to write down the analog of (\ref{diffuv}) in the infrared (under the assumption of supersymmetry enhancement) and use it to explicitly evaluate $D^{\rm IR}(x)$. Using then (\ref{uvir}) we can determine the scaling dimension of all the other operators.

From the RG independence of relative scaling dimensions we know that the ratio (which we denote by $\alpha$) between the dimension of $s_1$ and the dimension of $z$ is the same in the UV and in the IR, and indeed we know how to compute it in the UV: 
\be\label{scale3}
\alpha^{\rm IR}=\alpha^{\rm UV}=\frac{D^{\rm UV}(s_1)}{D^{\rm UV}(z)}\,.
\ee 
This relation is useful in writing down the infrared counterpart of (\ref{diffuv}): As we have already explained in Section \ref{sec:4d}, whenever supersymmetry enhancement occurs, $s_1$ is identified with the CB operator in the IR, and therefore (\ref{scaleuv}) should be replaced by 
\be\label{scaleir}
\frac{\partial\lambda_{\rm SW}}{\partial s_1}=\frac{dx}{y}\,,
\ee 
leading to the equation 
\be\label{scale4}
1-\alpha D^{\rm IR}(z)=-\frac{D^{\rm IR}(x)}{2}\,.
\ee 
Using now (\ref{uvir}) and (\ref{scale3}), we can rewrite (\ref{scale4}) as follows: 
\be\label{scaleir}
D^{\rm IR}(x)=\frac{2D^{\rm UV}(x)}{2D^{\rm UV}(s_1)-D^{\rm UV}(x)}\,,
\ee
expressing $D^{\rm IR}(x)$ in terms of UV scaling dimensions, as desired. Once $D^{\rm IR}(x)$ is known, we can determine all the other scaling dimensions from the curve (see (\ref{scale2})). In conclusion, this equation can be used to fix completely the scaling dimensions in the IR. At this stage it is rather simple to implement the consistency condition (\ref{unitarity}) for enhancement, namely the decoupling of $s_2$ and $z$:
\be
D^{\rm IR}(z)\leq1\,;\qquad D^{\rm IR}(s_2)\leq1\,.
\ee 
Using again the RG invariance of relative scaling dimensions (\ref{uvir}), these inequalities can be written more explicitly as follows:
\be\label{test} 
\frac{2D^{\rm UV}(z)}{2D^{\rm UV}(s_1)-D^{\rm UV}(x)}\leq1\,;\qquad \frac{2D^{\rm UV}(s_2)}{2D^{\rm UV}(s_1)-D^{\rm UV}(x)}\leq1\,.
\ee
If the choice of nilpotent orbit is not consistent with these inequalities, we conclude that supersymmetry does not enhance. In the rest of the Appendix we will check that (\ref{test}) singles out precisely the nilpotent vev's which induce enhancement of supersymmetry in the infrared for the theories $D_4$, $E_6$, $E_7$ and $E_8$.

\subsection{Flows starting from $D_4$}
In Table \ref{D4orbits} we list all the nilpotent orbits of $D_4$. The orbits colored in red are the ones in which the highest-dimensional irrep appears more than once as well as those such that the singlet belonging to the highest-spin irrep has in the UV a conformal dimension smaller or equal to $2$ (i.e. the dimension of the CB operator). As we have seen in Section \ref{sec:4d}, for these orbits there can never be enhancement. This leaves us with a total of 5 cases to be checked by hand.

\begin{table}[h!]
	\centering
	\begin{tabular}{|c|c|c|c|}
		\hline 
		Orbit $\mathcal{O}$ & $\dim_{\mathbb{C}}\bar{\mathcal{O}}$ & Decomposition of Adj &Enhancement? \\ 
		\hline
		\hline 
		$[7,1]$ & $24$ &$V_1\oplus 2V_3\oplus V_5$ & Yes, $\mathcal{H}_0$\\ 
		\hline 
		$[5,3]$ & $22$ &$\color{red} 3V_1\oplus V_2\oplus 2V_3$  & No\\ 
		\hline 
		$[5,1^{3}]$ & $20$ & $3V_0\oplus V_1\oplus 3V_2\oplus V_3$ & Yes, $\mathcal{H}_1$\\ 
		\hline 
		$[4^2]^{I}$ & $20$ & $3V_0\oplus V_1\oplus 3V_2\oplus V_3$ & Yes, $\mathcal{H}_1$\\ 
		\hline 
		$[4^2]^{II}$ & $20$ & $3V_0\oplus V_1\oplus 3V_2\oplus V_3$ & Yes, $\mathcal{H}_1$\\ 
		\hline 
		$[3^{2},1^{2}]$ & $18$ & $2V_0\oplus 7V_1\oplus V_2$ & Yes, $\mathcal{H}_2$\\ 
		\hline 
		$[3,2^{2},1]$ & $16$ & $\color{red} 3V_0\oplus 4V_{\frac{1}{2}}\oplus 3V_1\oplus 2V_{\frac{3}{2}}$ & No\\ 
		\hline 
		$[3,1^{5}]$ & $12$ & $\color{red} 10V_0\oplus 6V_1$ & No\\ 
		\hline 
		$[2^{4}]^{I}$ & $12$ & $\color{red} 10V_0\oplus 6V_1$ & No\\ 
		\hline 
		$[2^{4}]^{II}$ & $12$ & $\color{red} 10V_0\oplus 6V_1$ & No\\ 
		\hline 
		$[2^2,1^{4}]$ & $10$ & $\color{red} 9V_0\oplus 8V_{\frac{1}{2}}\oplus V_1$ & No\\ 
		\hline 
		$[1^{8}]$ & $0$ & $\color{red} 28V_0$ & No\\ 
		\hline 
	\end{tabular}
	\caption{All nilpotent orbits of $\mathfrak{so}(8)$ and their adjoint decomposition. By $V_j$ we indicate the representation of spin $j$ under the embedded $SU(2)$.} 
		\label{D4orbits}
\end{table}

Remarkably, all five cases will give enhancement. Therefore, simply by imposing the two criteria of uniqueness of the highest-spin singlet and that its UV dimension is greater than $2$, will single out all and only the enhancing orbits. There is in principle no need to use the inequalities (\ref{test}) in this case. Nevertheless, checking them, one sees that they are satisfied in the enhancing cases, as they should be. Let us perform this check explicitly.

The undeformed Weierstrass model reads (see Table \ref{table1})
\begin{equation}
y^2=x^3+ \tau x z^2 + z^3 
\end{equation}
From the homogeneity of the curve and the CY condition we can determine the dimensions of the coordinates and of the CB operator. We get
\begin{equation}
D^{\rm UV}(x)=2, \qquad D^{\rm UV}(y)=3, \qquad  D^{\rm UV}(z)=2.
\end{equation}

The inequalities (\ref{test}) now read
\begin{equation}
\begin{aligned}
D^{\rm UV}(s_1) &\geq 3\\
D^{\rm UV}(s_1)&\geq D^{\rm UV}(s_2)+1
\end{aligned}
\label{ineqD4}
\end{equation}
and simply by looking at Table \ref{D4orbits}, we can see that they are indeed satisfied in the enhancing cases.

\subsection{Flows starting from $E_6$}

The undeformed Weierstrass model in this case reads (see Table \ref{table1})
\begin{equation}
y^2=x^3+z^4\,.
\end{equation}
From the homogeneity of the curve and the CY condition we can determine the dimensions of the coordinates and of the CB operator. We get
\begin{equation}
	\begin{aligned}
	D^{\rm UV}(x)=4\,,\\
	D^{\rm UV}(y)=6\,,\\
	D^{\rm UV}(z)=3\,.
	\end{aligned}
\end{equation} 
In Table \ref{tableE6} we list all the nilpotent orbits of $\mathfrak{e}_6$. Again we color in red orbits with the highest-spin UV dimension less or equal to 3 (i.e. the UV dimension of the CB operator) or in which the highest-spin irrep appears more than once. As we have seen in Section \ref{sec:4d}, for these orbits there can never be enhancement. This leaves us with a total of $10$ cases to be still checked. 

The inequalities (\ref{test}) in this case read
\begin{equation}
\begin{aligned}
D^{\rm UV}(s_1) &\geq 5\,,\\
D^{\rm UV}(s_1)&\geq D^{\rm UV}(s_2)+2\,.
\end{aligned}
\label{ineqE6}
\end{equation}
In Table \ref{tableE6} we color in blue all the orbits which violate (\ref{ineqE6}), leaving just the orbits which give supersymmetry enhancement.

\begin{table}[h!]
	\centering
	\begin{tabular}{|c|c|c|c|}
		\hline
		Orbit $\mathcal{O}$ & $\dim_{\mathbb{C}}\bar{\mathcal{O}}$ & Decomposition of Adj &Enhancement? \\ 
		\hline
		\hline 
		$E_6$ & $72$ & $ V_1 \oplus V_4 \oplus V_5 \oplus V_7 \oplus V_8 \oplus V_{11}$ & Yes, $\mathcal{H}_0$ theory. \\ 
		\hline 
		$E_6(a_1)$ & $70$ & $\color{blue} V_1 \oplus V_2 \oplus V_3 \oplus V_4 \oplus 2V_5 \oplus V_7 \oplus V_8$ & No \\ 
		\hline 
		$D_5$ & $68$ & $ V_0 \oplus V_1 \oplus 2V_2 \oplus V_3 \oplus V_4 \oplus 3V_5 \oplus V_7$ & Yes, $\mathcal{H}_1$ theory. \\ 
		\hline 
		$E_6(a_3)$ & $66$ & $\color{red} 3V_1 \oplus 3V_2 \oplus 2V_3 \oplus 2V_4 \oplus 2V_5$ & No \\ 
		\hline 
		$D_5(a_1)$ & $64$ & $\color{blue} V_0 \oplus 2V_{\frac{1}{2}} \oplus 2V_1 \oplus V_2 \oplus2V_{\frac{5}{2}}\oplus 2V_3 \oplus 2V_{\frac{7}{2}}\oplus V_4 \oplus V_5$ & No \\ 
		\hline 
		$A_5$ & $64$ & $\color{blue} 3V_0 \oplus V_1\oplus 2V_{\frac{3}{2}}\oplus V_2 \oplus 2V_{\frac{5}{2}} \oplus V_3 \oplus V_4 \oplus 2V_{\frac{9}{2}} \oplus V_5$ & No \\ 
		\hline 
		$A_4+A_1$ & $62$ & $\color{blue} V_0 \oplus 2V_{\frac{1}{2}}\oplus 2V_1 \oplus 2V_{\frac{3}{2}}\oplus 3V_2 \oplus 2V_{\frac{5}{2}}\oplus V_3 \oplus 2V_{\frac{7}{2}}\oplus V_4$ & No \\ 
		\hline 
		$D_4$ & $60$ & $ 8V_0\oplus V_1 \oplus 8V_3 \oplus V_5$ & Yes, $\mathcal{H}_2$ theory. \\ 
		\hline 
		$A_4$ & $60$ & $\color{blue} 4V_0 \oplus 5V_1 \oplus 3V_2 \oplus 5V_3 \oplus V_4$ & No \\ 
		\hline 
		$D_4(a_1)$ & $58$ & $\color{red} 2V_0 \oplus 9V_1 \oplus 7V_2 \oplus 2V_3$ & No \\ 
		\hline 
		$A_3+A_1$ & $56$ & $\color{blue} 4V_0\oplus 2V_{\frac{1}{2}}\oplus 4V_1 \oplus 6V_{\frac{3}{2}}\oplus 3V_2 \oplus 2V_{\frac{5}{2}}\oplus V_3$ & No \\ 
		\hline 
		$2A_32+A_1$ & $54$ & $\color{red} 3V_0\oplus 6V_{\frac{1}{2}}\oplus  5V_1 \oplus 4V_{\frac{3}{2}} \oplus 4V_2 \oplus 2V_{\frac{5}{2}}$& No \\ 
		\hline 
		$A_3$ & $52$ & $\color{blue} 11V_0 \oplus V_1 \oplus 8V_{\frac{3}{2}} \oplus 5V_2 \oplus V_3$ & No \\ 
		\hline 
		$A_2+2A_1$ & $50$ & $\color{red} 4V_0\oplus 8V_{\frac{1}{2}} \oplus 9V_1 \oplus 4V_{\frac{3}{2}}\oplus 3V_2$& No \\ 
		\hline 
		$2A_2$ & $48$ & $\color{red} 14V_0 \oplus 8V_1\oplus 8V_2$ & No \\ 
		\hline 
		$A_2+A_1$ & $46$ & $\color{red} 9V_0\oplus 8V_{\frac{1}{2}} \oplus 8V_1 \oplus 6V_\frac{3}{2}\oplus V_2$ & No \\ 
		\hline 
		$A_2$ & $42$ & $\color{red}16 V_0\oplus 19 V_1\oplus V_2$ & No \\ 
		\hline 
		$3A_1$ & $40$ & $\color{red} 11V_0\oplus 16V_{\frac{1}{2}}\oplus 9V_1\oplus 2V_{\frac{3}{2}} $ & No \\ 
		\hline 
		$2A_1$ & $32$ & $\color{red} 22V_0 \oplus 16V_{\frac{1}{2}} \oplus 8V_1$ & No \\ 
		\hline 
		$A_1$ & $22$ & $\color{red} 35V_0\oplus  20V_{\frac{1}{2}}\oplus V_1$ & No \\ 
		\hline 
		$0$ & $0$ & $\color{red} 78V_0$ & No \\ 
		\hline 
	\end{tabular}
	\caption{All the nilpotent orbits of $\mathfrak{e}_6$ and their adjoint decomposition. Table taken from \cite{Hanany:2017ooe}.} 
	\label{tableE6}
\end{table}

\subsection{Flows starting from $E_7$}

The undeformed Weierstrass model in this case reads (see Table \ref{table1})
\begin{equation}
y^2=x^3+xz^3\,.
\end{equation}
From the homogeneity of the curve and the CY condition we get
\begin{equation}
D^{\rm UV}(x)=6\,, \qquad D^{\rm UV}(y)=9\,, \qquad D^{\rm UV}(z)=4\,.
\end{equation}

In Table \ref{tableE7} we list all the nilpotent orbits of $\mathfrak{e}_7$. The orbits colored in red are the ones in which the highest-dimensional irrep appears more than once. As we have seen, for these orbits there can never be enhancement. We also color in red the orbits for which the singlet $s_1$ belonging to the highest-spin irrep has in the UV a conformal dimension smaller than or equal to that of the CB operator, which is $4$. They also can never lead to enhancement, as discussed in Section \ref{sec:4d}. This leaves us with a total of $20$ cases to be checked by hand.

The consistency conditions (\ref{test}) in this case read
\begin{equation}
\begin{aligned}
D^{\rm UV}(s_1) &\geq 7\,,\\
D^{\rm UV}(s_1)&\geq D^{\rm UV}(s_2)+3\,.
\end{aligned}
\label{ineqE7}
\end{equation}
In Table (\ref{tableE7}) we color in blue all the orbits which violate (\ref{ineqE7}). It is trivial to see from the table that such inequalities are satisfied only by orbits which give supersymmetry enhancement.
\newpage

\begin{table}[h!]
	\centering
	\scalebox{0.65}{\begin{tabular}{|c|c|c|c|}
		\hline 
		Orbit $\mathcal{O}$ & $\dim_{\mathbb{C}}\bar{\mathcal{O}}$ & Decomposition of Adj &Enhancement? \\ 
		\hline
		\hline 
		$E_7$ & $126$ & $ V_1\oplus  V_5\oplus V_7\oplus V_9\oplus V_{11}\oplus V_{13}\oplus V_{17}$ & Yes, $\mathcal{H}_0$ theory. \\ 
		\hline 
		$E_7(a_1)$ & $124$ & $\color{blue} V_1\oplus V_3\oplus 2V_5\oplus V_7\oplus V_8\oplus V_9\oplus V_{11}\oplus V_{13}$ & No \\ 
		\hline 
		$E_7(a_2)$ & $122$ & $\color{blue} 2V_1\oplus V_3\oplus V_4\oplus 2V_5\oplus 2V_7\oplus V_8\oplus V_9\oplus V_{11}$ & No \\ 
		\hline 
		$E_7(a_3)$ & $120$ & $\color{blue} 2V_1\oplus V_2\oplus 2V_3\oplus V_4\oplus 4V_5\oplus 2V_7\oplus V_8\oplus V_9$ & No \\ 
		\hline 
		$E_6$ & $120$ & $ 3V_0\oplus V_1\oplus 3V_4\oplus V_5\oplus V_7\oplus 3V_8\oplus V_{11}$ & Yes, $\mathcal{H}_1$ theory. \\ 
		\hline 
		$E_6(a_1)$ & $118$ & $\color{blue} V_0\oplus V_1\oplus 3V_2\oplus V_3\oplus 3V_4\oplus 2V_5\oplus 2V_6\oplus V_7\oplus V_8$ & No \\ 
		\hline 
		$D_6$ & $118$ & $\color{blue} 3V_0\oplus V_1\oplus 2V_{\frac{5}{3}}\oplus V_3\oplus 2V_{\frac{9}{2}}\oplus 2V_5\oplus V_7\oplus 2V_{\frac{15}{2}}\oplus V_9$ & No \\ 
		\hline 
		$E_7(a_4)$ & $116$ & $\color{blue} 4V_1\oplus 2V_2\oplus 3V_3\oplus 2V_4\oplus 5V_5\oplus V_6\oplus V_7$ & No \\ 
		\hline 
		$D_6(a_1)$ & $114$ & $\color{blue} 3V_0\oplus 2V_1\oplus 2V_{\frac{3}{2}}\oplus 2V_{\frac{5}{2}}\oplus 2V_3\oplus V_4\oplus 2V_{\frac{9}{2}}\oplus 2V_5\oplus 2V_{\frac{11}{2}}\oplus V_7$ & No \\ 
		\hline 
		$D_5 + A_1$ & $114$ & $\color{blue} 3V_0\oplus 2V_1\oplus 2V_{\frac{3}{2}}\oplus 2V_{\frac{5}{2}}\oplus V_3\oplus 3V_4\oplus 2V_{\frac{9}{2}}\oplus V_5\oplus 2V_{\frac{11}{2}}\oplus V_7$ & No \\ 
		\hline 
		$A_6$ & $114$ & $\color{red} 3V_0\oplus V_1\oplus 3V_2\oplus 5V_3\oplus 3V_4\oplus V_5\oplus 3V_6$ & No \\ 
		\hline 
		$E_7(a_5)$ & $112$ & $\color{red} 6V_1\oplus 4V_2\oplus 5V_3\oplus 3V_4\oplus 3V_5$ & No \\ 
		\hline 
		$D_5$ & $112$ & $ \color{blue} 6V_0\oplus V_1\oplus 4V_2\oplus V_3\oplus 3V_4\oplus  5V_5\oplus V_7$ & No \\ 
		\hline 
		$E_6(a_3)$ & $110$ & $\color{red} 3V_0\oplus 3V_1\oplus 7V_2\oplus 4V_3\oplus 4V_4\oplus 2V_5$ & No \\ 
		\hline 
		$D_6(a_2)$ & $110$ & $\color{red} 3V_0\oplus 3V_1\oplus 4V_{\frac{3}{2}}\oplus V_2\oplus 2V_{\frac{5}{2}}\oplus 3V_3\oplus 2V_{\frac{7}{2}}\oplus V_4\oplus 2V_{\frac{9}{2}}\oplus 2V_5$ & No \\ 
		\hline 
		$D_5(a_1) + A_1$ & $108$ & $\color{blue} 3V_0\oplus 7V_1\oplus 3V_2\oplus 8V_3\oplus 3V_4\oplus V_5$ & No \\ 
		\hline 
		$A_5 + A_1$ & $108$ & $\color{blue} 3V_0\oplus 4V_{\frac{1}{2}}\oplus 2V_1\oplus 2V_{\frac{3}{2}}\oplus 3V_2\oplus 2V_{\frac{5}{2}}\oplus V_3\oplus 2V_{\frac{7}{2}}\oplus 3V_4\oplus 2V_{\frac{9}{2}}\oplus V_5$ & No \\ 
		\hline 
		$(A_5)'$ & $108$ & $\color{blue} 6V_0\oplus V_1\oplus 2V_{\frac{3}{2}}\oplus 3V_2\oplus 6V_{\frac{5}{2}}\oplus V_3\oplus 3V_4\oplus 2V_{\frac{9}{2}}\oplus V_5$ & No \\ 
		\hline 
		$A_4+A_2$ & $106$ & $\color{red} 3V_0\oplus 6V_1\oplus 10V_2\oplus 5V_3\oplus 3V_4$ & No \\  
		\hline 
		$D_5(a_1)$ & $106$ & $\color{blue} 4V_0\oplus 4V_{\frac{1}{2}}\oplus 4V_1\oplus V_2\oplus 4V_{\frac{5}{2}}\oplus 4V_3\oplus 4V_{\frac{7}{2}}\oplus V_4\oplus V_5$ & No \\  
		\hline 
		$A_4+A_1$ & $ 104$ & $\color{blue} 2V_0\oplus 4V_{\frac{1}{2}}\oplus 4V_1\oplus 4V_{\frac{3}{2}}\oplus 5V_2\oplus 4V_{\frac{5}{2}}\oplus 3V_3\oplus 2V_{\frac{7}{2}}\oplus V_4$ & No \\  
		\hline 
		$D_4+A_1$ & $102$ & $\color{blue} 10V_0\oplus 4V_{\frac{1}{2}}\oplus 4V_1\oplus 4V_{\frac{3}{2}}\oplus 5V_2\oplus 4V_{\frac{5}{2}}\oplus 3V_3\oplus 2V_{\frac{7}{2}}\oplus V_4$ & No \\  
		\hline 
		$(A_5)''$ & $102$ & $\color{blue} 14V_0\oplus V_1\oplus 7V_2\oplus V_3\oplus 7V_4\oplus V_5$ & No \\  
		\hline 
		$A_3+A_2+A_1$ & $100$ & $\color{red} 3V_0\oplus 15V_1\oplus 10V_2\oplus 15V_3$ & No \\  
		\hline 
		$A_4$ & $100$ & $\color{blue} 9V_0\oplus 7V_1\oplus 9V_2\oplus 7V_3\oplus V_4$ & No \\  
		\hline 
		$A_3+A_2$ & $98$ & $\color{red} 4V_0\oplus 4V_{\frac{1}{2}}\oplus 8V_1\oplus 8V_{\frac{3}{2}}\oplus 4V_2\oplus 4V_{\frac{5}{2}}\oplus 3V_3$ & No \\  
		\hline 
		$D_4(a_1)+A_1$ & $96$ & $\color{red} 6V_0\oplus 4V_{\frac{1}{2}}\oplus 8V_1\oplus 8V_{\frac{3}{2}}\oplus 5V_2\oplus 4V_{\frac{5}{2}}\oplus 2V_3$ & No \\  
		\hline 
		$D_4$ & $96$ & $\color{blue} 21V_0\oplus V_1\oplus 14V_3\oplus V_5$ & No \\  
		\hline 
		$A_3+2A_1$ & $94$ & $\color{red} 6V_0\oplus 8V_{\frac{1}{2}}\oplus 7V_1\oplus 6V_{\frac{3}{2}}\oplus 7V_2\oplus 4V_{\frac{5}{2}}\oplus V_3$ & No \\  
		\hline 
		$D_4(a_1)$ & $94$ & $\color{red} 9V_0\oplus 15V_1\oplus 13V_2\oplus 2V_3$ & No \\  
		\hline 
		$(A_3+A_1)'$ & $92$ & $\color{red}9V_0\oplus 6V_{\frac{1}{2}}\oplus 6V_1\oplus 10V_{\frac{3}{2}}\oplus 7V_2\oplus 2V_{\frac{5}{2}}\oplus V_3$ & No \\ 
		\hline 
		$2A_2+2A_1$ & $90$ & $\color{red} 6V_0\oplus 10V_{\frac{1}{2}}\oplus 11V_1\oplus 8V_{\frac{3}{2}}\oplus 6V_2\oplus 2V_{\frac{5}{2}}$ & No \\  
		\hline 
		$(A_3+A_1)'$ & $86$ & $\color{red}21V_0\oplus 10V_1\oplus 15V_2\oplus V_3$ & No \\ 
		\hline 
		$A_2+ 3A_1$ & $84$ & $\color{red} 14V_0\oplus 28V_1\oplus 7V_2$ & No \\  
		\hline 
		$2A_2$ & $84$ & $\color{red} 17V_0\oplus 22V_1\oplus 10V_2$ & No \\  
		\hline 
		$A_3$ & $84$ & $\color{red} 24V_0\oplus V_1\oplus 16V_{\frac{3}{2}}\oplus 7V_2\oplus V_3$ & No \\  
		\hline 
		$A_2+2A_1$ & $82$ & $\color{red} 9V_0\oplus 16V_{\frac{1}{2}}\oplus 15V_1\oplus 8V_{\frac{3}{2}}\oplus 3V_2$ & No \\  
		\hline 
		$A_2+A_1$ & $76$ & $\color{red}16V_0\oplus 16V_{\frac{1}{2}}\oplus 16V_1\oplus 8V_{\frac{3}{2}}\oplus V_2$ & No \\  
		\hline 
		$4A_1$ & $70$ & $\color{red}21V_0\oplus 20V_{\frac{1}{2}}\oplus 16V_1\oplus 6V_{\frac{3}{2}}$ & No \\  
		\hline 
		$A_2$ & $66$ & $\color{red} 35V_0\oplus 31V_1\oplus V_2$ & No \\  
		\hline 
		$(3A_1)'$ & $64$ & $\color{red} 24V_0\oplus 28V_{\frac{1}{2}}\oplus 2V_{\frac{3}{2}}$ & No \\  
		\hline 
		$(3A_1)'$ & $54$ & $\color{red} 52V_0\oplus 27V_1$ & No \\  
		\hline 
		$2A_1$ & $52$ & $\color{red} 39V_0\oplus 32V_{\frac{1}{2}}\oplus 10V_1$ & No \\  
		\hline 
		$A_1$ & $34$ & $\color{red} 66V_0\oplus 32V_{\frac{1}{2}}\oplus V_1$ & No \\  
		\hline 
		$0$ & $0$ & $\color{red} 133V_0$ & No \\  
		\hline 
	\end{tabular} }
	\caption{All the nilpotent orbits of $\mathfrak{e}_7$ and their adjoint decomposition. Table taken from \cite{Hanany:2017ooe}.}
	\label{tableE7}
\end{table}

\newpage

\subsection{Flows starting from $E_8$}

From the UV surface (see Table \ref{table1})
\begin{equation}
y^2=x^3+z^5\,,
\end{equation}
we find the following assignment of scaling dimensions:
\begin{equation}
\begin{aligned}
D^{\rm UV}(x)=10\,,\\
D^{\rm UV}(y)=15\,,\\
D^{\rm UV}(z)=6\,.
\end{aligned}
\end{equation}

In Table (\ref{tableE8p1}) we list all the nilpotent orbits of $\mathfrak{e}_8$. Again we have colored in red all the orbits which cannot induce enhancement of supersymmetry, either because the highest-dimensional irrep appears more than once, or because the singlet belonging to the highest-spin irrep has in the UV a conformal dimension smaller than or equal to that of the CB operator. Since in this case the CB operator has dimension $6$, the latter condition rules out all the orbits with highest spin smaller or equal to $5$. This leaves us with a total of $25$ cases to be checked by hand.

We now have to impose the unitarity constraints (\ref{test}), which reads
\begin{equation}
\begin{aligned}
D^{\rm UV}(s_1) &\geq 11\,,\\
D^{\rm UV}(s_1)&\geq D^{\rm UV}(s_2)+5\,.
\end{aligned}
\label{ineqE8}
\end{equation}
In Table (\ref{tableE8p1}) we color in blue all the orbits which violate inequalities (\ref{ineqE8}). It is easy to see from the table that such inequalities are satisfied only by the principal nilpotent orbit, which therefore is the only one leading to IR supersymmetry enhancement.
\newpage

\begin{table}[h!]
	\centering
		\scalebox{0.57}{\begin{tabular}{|c|c|c|c|}
		\hline 
		Orbit $\mathcal{O}$ & $\dim_{\mathbb{C}}\bar{\mathcal{O}}$ & \thead{Decomposition of Adj} &Enhancement? \\ 
		\hline \hline 
		$E_8$ & $240$ & $ V_1\oplus V_7 \oplus V_{11}\oplus V_{13}\oplus V_{17}\oplus V_{19}\oplus V_{23}\oplus V_{29}$ & Yes, $\mathcal{H}_0$ theory \\ 
		\hline 
		$E_8(a_1)$ & $238$ & $\color{blue}V_1\oplus V_5 \oplus V_7 \oplus V_9 \oplus V_{11} \oplus V_{13} \oplus V_{14} \oplus V_{17} \oplus V_{19} \oplus V_{23}$ & No \\ 
		\hline 
		$E_8(a_2)$ & $236$ & $\color{blue}V_1 \oplus V_3 \oplus V_5 \oplus V_7 \oplus V_8 \oplus V_9 \oplus 2V_{11} \oplus V_{13}\oplus V_{14} \oplus V_{17} \oplus V_{19}$ & No \\ 
		\hline 
		$E_8(a_3)$ & $234$ & $\color{blue}2V_1\oplus V_4 \oplus 2V_5 \oplus V_7 \oplus V_8 \oplus 2V_9 \oplus V_{11} \oplus 2V_{13}\oplus V_{14} \oplus V_{17}$ & No \\ 
		\hline 
		$E_8(a_4)$ & $232$ & $\color{blue}V_1\oplus V_2\oplus V_3\oplus V_4\oplus 2V_5\oplus 3V_7\oplus V_8\oplus 2V_9\oplus 2V_{11}\oplus V_{13}\oplus V_{14}$ & No \\ 
		\hline 
		$E_7$ & $232$ & $\color{blue}3V_0\oplus V_1\oplus 2V_{\frac{9}{2}}\oplus V_5\oplus V_7\oplus 2V_{\frac{17}{2}}\oplus V_9\oplus V_{11}\oplus V_{13}\oplus 2V_{\frac{27}{2}}\oplus V_{17}$ & No \\ 
		\hline 
		$E_8(b_4)$ & $230$ & $\color{blue} 2V_1\oplus V_2\oplus 2V_3\oplus 2V_5\oplus V_6\oplus 2V_7\oplus 2V_8\oplus V_9\oplus V_{10}\oplus 2V_{11}\oplus V_{13}$ & No \\ 
		\hline 
		$E_8(a_5)$ & $228$ & $\color{red} 3V_1\oplus V_2\oplus V_3\oplus V_4\oplus 4V_5\oplus 2V_6\oplus 3V_7\oplus V_8\oplus V_9\oplus V_{10}\oplus 2V_{11}$ & No \\ 
		\hline 
		$E_7(a_1)$ & $228$ & \color{blue}\makecell{$3V_0\oplus V_1\oplus 2V_{\frac{5}{2}}\oplus V_3 \oplus 2V_5\oplus 2V_{\frac{11}{2}}\oplus V_7\oplus$\\$\oplus 2V_{\frac{15}{2}}\oplus V_8\oplus V_9\oplus 2V_{\frac{21}{2}}\oplus V_{11}\oplus V_{13}$} & No \\ 
		\hline 
		$E_8(b_5)$ & $226$ & \color{blue}$4V_1\oplus V_2\oplus 2V_3\oplus 3V_4\oplus 3V_5\oplus 3V_7\oplus 3V_8 \oplus 2V_9\oplus V_{11}$ & No \\ 
		\hline 
		$D_7$ & $226$ & \color{blue}\makecell{$3V_0\oplus V_1 \oplus 2V_{\frac{3}{2}}\oplus V_3 \oplus 2V_{\frac{9}{2}}\oplus V_5 \oplus 2V_{\frac{11}{2}}\oplus$\\$\oplus 3V_6\oplus V_7\oplus 2V_{\frac{15}{2}}\oplus V_9\oplus \oplus 2V_{\frac{21}{2}}\oplus V_{11}$} & No \\ 
		\hline 
		$E_8(a_6)$ & $224$ & $\color{red}3V_1\oplus V_2 \oplus 5V_3\oplus 3V_4\oplus 3V_5 \oplus 3V_6 \oplus 3V_7 \oplus V_8 \oplus 2V_9$ & No \\ 
		\hline 
		$E_7(a_2)$ & $224$ & \color{blue}\makecell{$3V_1\oplus 2V_2\oplus 2V_{\frac{3}{2}}\oplus V_3\oplus 2V_{\frac{7}{2}}\oplus V_4\oplus 2V_{\frac{9}{2}}\oplus 2V_5\oplus$\\$\oplus 2V_7\oplus 2V_{\frac{15}{2}}\oplus V_8\oplus 2V_{\frac{17}{2}}\oplus V_9\oplus V_{11}$} & No \\ 
		\hline 
		$E_6+A_1$ & $222$ & \color{blue}\makecell{$3V_0\oplus 4V_{\frac{1}{2}}\oplus 2V_1\oplus 2V_{\frac{7}{2}}\oplus 3V_4\oplus 2V_{\frac{9}{2}}\oplus$\\$\oplus V_5\oplus V_7\oplus 2V_{\frac{15}{2}}\oplus 3V_8\oplus 2V_{\frac{17}{2}}\oplus V_{11}$} & No \\ 
		\hline 
		$D_7(a_1)$ & $222$ & $\color{blue}V_0\oplus 4V_1 \oplus 2V_2 \oplus 3V_3 \oplus 3V_4 \oplus 6V_5 \oplus V_6 \oplus 3V_7 \oplus 2V_8 \oplus V_9$ & No \\ 
		\hline 
		$E_8(b_6)$ & $220$ & $\color{blue}4V_1 \oplus 4V_2 \oplus 5\oplus_3 \oplus 3V_4 \oplus 6V_5 \oplus 2V_6 \oplus 3V_7 \oplus V_8$ & No \\ 
		\hline 
		$E_7(a_3)$ & $220$ & \color{blue}\makecell{$3V_0\oplus 2V_{\frac{1}{2}}\oplus 2V_1 \oplus V_2 \oplus 2V_{\frac{5}{2}}\oplus 2V_3 \oplus V_4\oplus$\\$\oplus 4V_{\frac{9}{2}}\oplus 3V_{5} \oplus 2V_{\frac{11}{2}}\oplus 2V_7 \oplus 2V_{\frac{15}{2}}\oplus V_8 \oplus V_9 $} & No \\ 
		\hline 
		$E_6(a_1)+A_1$ & $218$ & \color{blue}\makecell{$V_0 \oplus 2V_{\frac{1}{2}}\oplus 2V_{1}\oplus 2V_{\frac{3}{2}}\oplus 3V_2 \oplus 2V_{\frac{5}{2}}\oplus V_3 \oplus 2V_{\frac{7}{2}}\oplus$\\$ \oplus 3V_4 \oplus 2V_{\frac{9}{2}} \oplus 2V_5\oplus 2V_{\frac{11}{2}}\oplus 2V_6 \oplus 2V_{\frac{13}{2}}\oplus V_7\oplus V_8$}& No \\ 
		\hline 
		$A_7$ & $218$ & \color{red}\makecell{$3V_0\oplus V_1 \oplus 2V_{\frac{3}{2}}\oplus 3V_2\oplus 2V_{\frac{5}{2}}\oplus V_3\oplus 4V_{\frac{7}{2}} \oplus$ \\$\oplus 3V_4\oplus 2V_{\frac{9}{2}}\oplus V_5\oplus 2V_{\frac{11}{2}}\oplus 3V_6 \oplus V_7 \oplus 2V_{\frac{15}{2}}$} & No \\ 
		\hline 
		$D_7(a_2)$ & $216$ & \color{blue}\makecell{$V_0 \oplus 2V_{\frac{1}{2}}\oplus 2V_1 \oplus 2V_{\frac{3}{2}}\oplus 3V_2 \oplus 2V_{\frac{5}{2}} \oplus 3V_3 \oplus 4V_{\frac{7}{2}}\oplus$\\$\oplus 3V_4 \oplus 2V_{\frac{9}{2}}\oplus 2V_5 \oplus 2V_{\frac{11}{2}}\oplus V_6 \oplus 2V_{\frac{13}{2}}\oplus V_7$} & No \\ 
		\hline 
		$E_6$ & $216$ & $\color{blue} 14V_0 \oplus V_1 \oplus 7V_4 \oplus V_5\oplus V_7 \oplus 7V_8 \oplus V_{11}$ & No \\ 
		\hline 
		$D_6$ & $216$ & $\color{blue} 10V_0 \oplus V_1 \oplus 4V_{\frac{5}{2}}\oplus V_3 \oplus 4V_{\frac{9}{2}}\oplus 6V_5 \oplus V_7 \oplus 4V_{\frac{15}{2}}\oplus V_9$ & No \\ 
		\hline 
		$D_5+A_2$ & $214$ & $\color{blue} V_0 \oplus 8V_1 \oplus 5V_2 \oplus 5V_3 \oplus 5V_4 \oplus 7V_5 \oplus 2V_6 \oplus V_7 $ & No \\ 
		\hline 
		$E_6(a_1)$ & $214$ & $\color{blue} 8V_0 \oplus V_1 \oplus 7V_2 \oplus V_3 \oplus 7V_4 \oplus 2V_5 \oplus 6V_6\oplus V_7 \oplus V_8 $ & No \\ 
		\hline 
		$E_7(a_4)$ & $212$ & \color{blue} \makecell{$3V_0 \oplus 2V_{\frac{1}{2}}\oplus 4V_1 \oplus 4V_{\frac{3}{2}}\oplus 2V_2 \oplus 2V_{\frac{5}{2}}\oplus 3V_3 \oplus 2V_{\frac{7}{2}}\oplus$\\$\oplus 2V_4 \oplus 4V_{\frac{9}{2}}\oplus 4V_5 \oplus 2V\frac{11}{2}\oplus V_6 \oplus V_7$} & No \\ 
		\hline 
		$A_6+A_1$ & $212$ & \color{red}\makecell{$3V_0 \oplus 2V_{\frac{1}{2}}\oplus 2V_1\oplus 2V_{\frac{3}{2}}\oplus 3V_2 \oplus 4V_{\frac{5}{2}}\oplus 5V_3 \oplus$\\$4V_{\oplus\frac{7}{2}}\oplus 3V_4 \oplus 2V\frac{9}{2}\oplus V_5 \oplus 2V_{\frac{11}{2}}\oplus 3V_6$} & No \\ 
		\hline 
		$D_6(a_1)$ & $210$ & $\color{blue} 6V_0 \oplus 6V_1 \oplus 4V_{\frac{3}{2}}\oplus 4V_{\frac{5}{2}}\oplus 2V_3 \oplus 5V_4 \oplus 4V_{\frac{9}{2}}\oplus 2V_5\oplus 4V_{\frac{11}{2}}\oplus V_7$ & No \\ 
		\hline 
		$A_6$ & $210$ & $\color{red}6V_0 \oplus 5V_1 \oplus 3V_2 \oplus 13 V_3 \oplus 3V_4 \oplus 5V_5 \oplus 3V_6$ & No \\ 
		\hline 
		$E_8(a_7)$ & $208$ & $\color{red}10V_1 \oplus 10V_2 \oplus 10V_3 \oplus 6V_4 \oplus 4V_5$ & No \\ 
		\hline 
		$D_5+A_1$ & $208$ & \color{blue} \makecell{$6V_0 \oplus 6V_{\frac{1}{2}}\oplus 2V_1 \oplus 2V_{\frac{3}{2}}\oplus 4V_2 \oplus 2V_{\frac{5}{2}}\oplus V_3\oplus$\\$\oplus 2V_{\frac{7}{2}}\oplus 3V_4 \oplus 4V_{\frac{9}{2}}\oplus 5V_5 \oplus 2V_{\frac{11}{2}}\oplus V_7$} & No \\ 
		\hline 
		$E_7(a_5)$ & $206$ & $\color{red}3V_0 \oplus 6V_1 \oplus 6V_{\frac{3}{2}}\oplus 4V_2 \oplus 6V_{\frac{5}{2}}\oplus 5V_3 \oplus 4V_{\frac{7}{2}}\oplus 3V_4\oplus 2V_{\frac{9}{2}}\oplus 3V_5$ & No \\ 
		\hline 
		$E_6(a_3)+A_1$ & $204$ &\color{red} \makecell{$3V_0 \oplus 4V_{\frac{1}{2}}\oplus 4V_1 \oplus 4V_{\frac{3}{2}}\oplus 7V_2 \oplus 6V_{\frac{5}{2}}\oplus$\\$\oplus 4V_3 \oplus 5V_{\frac{7}{2}}\oplus 4V_4 \oplus 2V_{\frac{9}{2}}\oplus 2V_5$} & No \\ 
		\hline 
		$D_6(a_2)$ & $204$ & $\color{red}6V_0\oplus 3V_1\oplus 8V_{\frac{3}{2}}\oplus 5V_2 \oplus 4V_{\frac{5}{2}}\oplus 7V_3 \oplus 4V_{\frac{7}{2}}\oplus V_4 \oplus 4V_{\frac{9}{2}}\oplus 2V_5$ & No \\ 
		\hline 
		$D_5(a_1)+A_2$ & $202$ & \color{red}\makecell{$3V_0 \oplus 4V_{\frac{1}{2}}\oplus 7V_1 \oplus 4V_{\frac{3}{2}}\oplus 6V_2 \oplus 6V_{\frac{5}{2}}\oplus$\\$\oplus 4V_3 \oplus 6V_{\frac{7}{2}}\oplus 3V_4 \oplus 2V_{\frac{9}{2}}\oplus V_5$} & No \\ 
		\hline 
		$A_5+A_1$ & $202$ & \color{red}\makecell{$6V_0\oplus 4V_{\frac{1}{2}}\oplus 2V_1 \oplus 4V_{\frac{3}{2}}\oplus 7V_2\oplus$\\$ \oplus 8V_{\frac{5}{2}}\oplus 5V_3 \oplus 2V_{\frac{7}{2}}\oplus 3V_4 \oplus 4V_{\frac{9}{2}}\oplus V_5$} & No \\ 
		\hline 
		$A_4+A_3$ & $200$ & $\color{red} 3V_0\oplus 4V_{\frac{1}{2}}\oplus 6V_1 \oplus 8V_{\frac{3}{2}}\oplus 6V_2 \oplus 6V_{\frac{5}{2}}\oplus 6V_3\oplus 4V_{\frac{7}{2}}\oplus 3V_4 \oplus 2V_{\frac{9}{2}}$ & No \\ 
		\hline 
		$D_5$ & $200$ & $\color{blue} 21V_0\oplus V_1 \oplus 8V_2 \oplus V_3 \oplus 7V_4 \oplus 9V_5 \oplus V_7$ & No \\ 
		\hline 
	\end{tabular} }
	\caption{All the orbits or $\mathfrak{e}_8$ and their adjoint decomposition. Part 1. Table taken from \cite{Hanany:2017ooe}.}
	\label{tableE8p1}
\end{table}

\begin{table}[t]	
	\centering
		\scalebox{0.57}{\begin{tabular}{|c|c|c|c|}
		\hline 
		Orbit $\mathcal{O}$ & $\dim_{\mathbb{C}}\bar{\mathcal{O}}$ & Decomposition of Adj & Enhancement? \\ 
		\hline 
		$E_6(a_3)$ & $198$ & $\color{red} 14V_0 \oplus 3V_1 \oplus 15V_2 \oplus 8V_3 \oplus 8V_4 \oplus 2V_5$ & No \\ 
		\hline 
		$D_4+A_2$ & $198$ & $\color{red}8V_0 \oplus 14 V_1 \oplus 7V_2 \oplus 14V_3 \oplus 6V_4 \oplus V_4$ & No \\ 
		\hline 
		$A_4+A_2+A_1$ & $196$ & $\color{red} 3V_0\oplus 6V_{\frac{1}{2}}\oplus 7V_1 \oplus 8V_{\frac{3}{2}}\oplus 10V_2 \oplus 6V_{\frac{5}{2}}\oplus 5V_3 \oplus 4V_{\frac{7}{2}}\oplus 3V_4$ & No \\ 
		\hline 
		$D_5(a_1)+A_1$ & $196$ & $\color{red}6V_0 \oplus 10V_{\frac{1}{2}}\oplus 7V_1 \oplus 2V_{\frac{3}{2}}\oplus 3V_2 \oplus 6V_{\frac{5}{2}}\oplus 8V_3 \oplus 6V_{\frac{7}{2}}\oplus 3V_4 \oplus V_5$ & No \\ 
		\hline 
		$A_5$ & $196$ & $\color{red}17V_0 \oplus V_1 \oplus 2V_{\frac{3}{2}}\oplus 7V_2 \oplus 14V_{\frac{5}{2}}\oplus V_3 \oplus 7V_4 \oplus 2V_{\frac{9}{2}}\oplus V_5$ & No \\ 
		\hline 
		$A_4+A_2$ & $194$ & $\color{red}6V_0 \oplus 18V_1 \oplus 14V_2 \oplus 13V_3 \oplus 3V_4$ & No \\ 
		\hline 
		$A_4+2A_1$ & $192$ & $\color{red}4 V_{0}\oplus 8V_{\frac{1}{2}}\oplus 9V_1 \oplus 8V_{\frac{3}{2}}\oplus 9V_2\oplus 8V_{\frac{5}{2}}\oplus 5V_3\oplus 4V\frac{7}{2}\oplus V_4$ & No \\ 
		\hline 
		$D_5(a_1)$ & $190$ & $\color{red}15 V_{0}\oplus 8V_{\frac{1}{2}}\oplus 8V_1 \oplus V_2\oplus 8V_{\frac{5}{2}}\oplus 8V_3\oplus 8V\frac{7}{2}\oplus V_4 \oplus V_5$ & No \\ 
		\hline 
		$2A_3$ & $188$ & $\color{red} 10V_0 \oplus 4V_{\frac{1}{2}}\oplus6V_1 \oplus 16V_{\frac{3}{2}}\oplus 10V_2 \oplus 4V_{\frac{5}{2}}\oplus 6V_3 \oplus 4V_{\frac{7}{2}}$ & No \\ 
		\hline 
		$A_4+A_1$ & $188$ & $\color{red}9V_0 \oplus 8V_{\frac{1}{2}}\oplus 8V_1 \oplus 8V_{\frac{3}{2}}\oplus 9V_{2}\oplus 8V_{\frac{5}{2}}\oplus 7V_3 \oplus 2V_{\frac{7}{2}}\oplus V_4$ & No \\ 
		\hline 
		$D_4(a_1)+A_2$ & $184$ & $\color{red} 8V_0 \oplus 28V_1 \oplus 20V_2 \oplus 8V_3$ & No \\ 
		\hline 
		$D_4+A_1$ & $184$ & $\color{red}21V_0\oplus 14V_{\frac{1}{2}} \oplus 2V_1\oplus 6V_{\frac{5}{2}}\oplus 14V_3 \oplus 6V_{\frac{7}{2}}\oplus V_5$ & No \\ 
		\hline 
		$A_3+A_2+A_1$ & $182$ & $\color{red} 6 V_{0}\oplus 10V_{\frac{1}{2}}\oplus 15V_1 \oplus 14V_{\frac{3}{2}}\oplus 10V_2\oplus 6V_{\frac{5}{2}}\oplus 5V_3$ & No \\ 
		\hline 
		$A_4$ & $180$ & $\color{red}24V_0\oplus 11V_1\oplus 21V_2\oplus  11V_3\oplus V_4$ & No \\ 
		\hline 
		$A_3+A_2$ & $178$ & $\color{red} 11 V_{0}\oplus 8V_{\frac{1}{2}}\oplus 16V_1 \oplus 16V_{\frac{3}{2}}\oplus 8V_2\oplus 8V_{\frac{5}{2}}\oplus 3V_3$ & No \\ 
		\hline 
		$D_4(a_1)+A_1$ & $176$ & $\color{red} 9 V_{0}\oplus 14V_{\frac{1}{2}}\oplus 16V_1 \oplus 12V_{\frac{3}{2}}\oplus 13V_2\oplus 6V_{\frac{5}{2}}\oplus 2V_3$ & No \\ 
		\hline 
		$A_3+2A_1$ & $172$ & $\color{red}13 V_{0}\oplus 14V_{\frac{1}{2}}\oplus 15V_1 \oplus 16V_{\frac{3}{2}}\oplus 11V_2\oplus 6V_{\frac{5}{2}}\oplus V_3$ & No \\ 
		\hline 
		$2A_2+2A_1$ & $168$ & $\color{red} 10 V_{0}\oplus 20V_{\frac{1}{2}}\oplus 20V_1 \oplus 16V_{\frac{3}{2}}\oplus 10V_2\oplus 4V_{\frac{5}{2}}$ & No \\ 
		\hline 
		$D_4$ & $168$ & $\color{red}52V_0 \oplus V_1\oplus 26V_3 \oplus V_5$ & No \\ 
		\hline
		$D_4(a_1)$ & $166$ & $\color{red} 28V_0\oplus 27V_1\oplus 25V_2\oplus 2V_3$ & No \\ 
		\hline 
		$A_3+A_1$ & $164$ & $\color{red}24 V_{0}\oplus 14V_{\frac{1}{2}}\oplus 10V_1 \oplus 18V_{\frac{3}{2}}\oplus 15V_2\oplus 2V_{\frac{5}{2}}\oplus V_3$ & No \\ 
		\hline 
		$2A_2+A_1$ & $162$ & $\color{red} 17 V_{0}\oplus 18V_{\frac{1}{2}}\oplus 23V_1 \oplus 16V_{\frac{3}{2}}\oplus 10V_2\oplus 2V_{\frac{5}{2}}$ & No \\ 
		\hline 
		$2A_2$ & $156$ & $\color{red} 28V_1\oplus 50V_3\oplus 14V_5$ & No \\ 
		\hline 
		$A_2+3A_1$ & $154$ & $\color{red} 17 V_{0}\oplus 28V_{\frac{1}{2}}\oplus 28V_1 \oplus 14V_{\frac{3}{2}}\oplus 7V_2$ & No \\ 
		\hline 
		$A_3$ & $148$ & $\color{red}55 V_{0}\oplus V_1 \oplus 32V_{\frac{3}{2}}\oplus 11V_2\oplus V_3$ & No \\ 
		\hline 
		$A_2+2A_1$ & $146$ & $\color{red}24 V_{0}\oplus 32V_{\frac{1}{2}}\oplus 27V_1 \oplus 16V_{\frac{3}{2}}\oplus 3V_2$ & No \\ 
		\hline 
		$A_2+A_1$ & $136$ & $\color{red}35 V_{0}\oplus 32V_{\frac{1}{2}}\oplus 32V_1 \oplus 14V_{\frac{3}{2}}\oplus V_2$ & No \\ 
		\hline 
		$4A_1$ & $128$ & $\color{red}36 V_{0}\oplus 48V_{\frac{1}{2}}\oplus 28V_1 \oplus 8V_{\frac{3}{2}}$ & No \\ 
		\hline 
		$A_2$ & $114$ & $\color{red} 78V_0 \oplus 55V_1\oplus V_2$ & No \\ 
		\hline 
		$3A_1$ & $112$ & $\color{red}55 V_{0}\oplus 52V_{\frac{1}{2}}\oplus 37V_1 \oplus 2V_{\frac{3}{2}}$ & No \\ 
		\hline 
		$2A_1$ & $92$ & $\color{red}78 V_{0}\oplus 64V_{\frac{1}{2}}\oplus 13V_1$ & No \\ 
		\hline 
		$A_1$ & $58$ & $\color{red} 133 V_{0}\oplus 56V_{\frac{1}{2}}\oplus V_1$ & No \\ 
		\hline 
		$0$ & $0$ & $\color{red} 248 V_0$ & No \\ 
		\hline 
	\end{tabular} }
	\caption{All the orbits or $\mathfrak{e}_8$ and their adjoint decomposition. Part 2. Table taken from \cite{Hanany:2017ooe}.}
	\label{tableE8p2}
\end{table}

\bibliographystyle{JHEP}
\bibliography{MSTrefs.bib}

\end{document}